\documentclass[trackchanges,twocolumn]{aastex701}

\usepackage{xspace}
\usepackage[nolist]{acronym}
\usepackage{graphicx}
\usepackage{subfigure}
\usepackage{xcolor}
\providecommand{\acrolowercase}[1]{\lowercase{#1}}

\usepackage{amsmath}

\begin{acronym}
\acro{KDE}[KDE]{kernel density estimation}
\acro{BSN}[BSN]{signal-to-noise coherent Bayes Factor}
\acro{BCI}[BCI]{the coherent versus incoherent Bayes Factor}
\acro{TPR}[TPR]{True Positive Rate}
\acro{FPR}[FPR]{False Positive Rate}
\acro{BNS}[BNS]{binary neutron star}
\acro{NSBH}[NSBH]{neutron star-black hole}
\acro{MDC}[MDC]{Mock Data Challenge}
\acro{EoS}[EoS]{equation of state}
\acro{NS}[NS]{neutron star}
\acro{GW}[GW]{gravitational-wave}
\acro{LIGO}[LIGO]{Laser Interferometer \acs{GW} Observatory}
\acro{aLIGO}[aLIGO]{Advanced \acs{LIGO}}
\acro{AdVirgo}[AdVirgo]{Advanced Virgo}
\acro{O4}[O4]{\ac{aLIGO}'s \citep{aLIGO}, \ac{AdVirgo}'s \citep{adVirgo}, and \ac{KAGRA}'s \citep{KAGRA:2020tym} fourth observing run}
\acro{O4a}[O4a]{first part of \ac{O4}}
\acro{LVK}[LVK]{LIGO-Virgo-KAGRA}
\acro{IGWN}[IGWN]{International Gravitational-Wave Network}

\acro{O1}[O1]{\ac{aLIGO}'s first observing run}
\acro{O2}[O2]{\ac{aLIGO}'s and \ac{AdVirgo}'s second observing run}
\acro{O3}[O3]{\ac{aLIGO}'s, \ac{AdVirgo}'s, and \ac{KAGRA}'s third observing run}

\acro{O5}[O5]{\ac{aLIGO}'s, \ac{AdVirgo}'s, and \ac{KAGRA}'s fifth observing run}
\acro{oLIB}[\acrolowercase{o}LIB]{Omicron+\acl{LIB}}
\acro{KAGRA}[KAGRA]{KAmioka GRAvitational\nobreakdashes-wave observatory}
\acro{GraceDB}[\texttt{GraceDB}]{GRAvitational-wave Candidate Event DataBase}
\acro{LLAI}[LLAI]{Low-Latency Alert Infrastructure}
\acro{CBC}[CBC]{compact binary coalescence}
\acro{BBH}[BBH]{binary black hole}
\acro{FAR}[FAR]{false alarm rate}
\acro{GstLAL}[GstLAL]{GStreamer LIGO Scientific Collaboration Algorithm Library}
\acro{SVD}[SVD]{singular value decomposition}
\acro{SNR}[SNR]{signal\nobreakdashes-to\nobreakdashes-noise ratio}
\acro{FGMC}[FGMC]{Farr, Gair, Mandel and Cutler formalism of Poisson Mixture Model}
\acro{GRB}[GRB]{Gamma Ray Burst}
\acro{PyCBC}[PyCBC]{Python based Compact Binary Coalescence search software}
\acro{MBTA}[MBTA]{Multi-Band Template Analysis}
\acro{SPIIR}[SPIIR]{Summed Parallel Infinite Impulse Response}
\acro{GPU}[GPU]{Graphical Processing Units}
\acro{cWB}[\acrolowercase{c}WB]{Coherent WaveBurst}
\acro{GCN}[GCN]{General Coordinates Network}
\acro{SCiMMA}[SCiMMA ]{Scalable CyberInfrastructure for Multi-Messenger Astrophysics}
\acro{ML}[ML]{Machine Learning}

\acro{MMA}[MMA]{Multi-Messenger Astronomy}
\acro{BAYESTAR}[\texttt{BAYESTAR}]{BAYESian TriAngulation and Rapid localization}
\acro{FITS}[FITS]{Flexible Image Transport System}
\acro{HEALPix}[HEALP\acrolowercase{ix}]{Hierarchical Equal Area isoLatitude Pixelization}
\acro{MOC}[MOC]{Multi-Order Coverage}
\acro{EM}[EM]{electromagnetic}
\acro{NS}[NS]{neutron star}
\acro{BH}[BH]{black hole}
\acro{LLOID}[LLOID]{Low Latency Online Inspiral Detection}
\acro{IIR}[IIR]{infinite impulse response}
\acro{ToO}[ToO]{target of opportunity}
\acro{OPA}[OPA]{Open Public Alert}
\acro{PE}[PE]{Parameter Estimation}
\acro{Bilby}[Bilby]{Bayesian Inference Library for \acs{CBC} \acs{GW} signal}
\acro{RAVEN}[RAVEN]{Rapid, on-source VOEvent Coincident Monitor}
\acro{KNN}[KNN]{K-Nearest Neighbors}
\acro{NN}[NN]{Neural Network}
\acro{MLP}[MLP]{Multilayer Perceptron}
\acro{RF}[RF]{Random Forest}
\acro{GWTC}[GWTC]{Gravitational-Wave Transient Catalog}
\acro{ROC}[ROC]{Receiver Operating Characteristic}
\acro{AUC}[AUC]{Area Under the Receiver Operating Characteristic Curve}
\acro{PDF}[PDF]{probability density function}
\acro{ZTF}[ZTF]{Zwicky Transient Facility}
\acro{LANL}[LANL]{Los Alamos National Laboratory}
\acro{PSF}[PSF]{point-spread function}
\end{acronym}
\begin{document}

\title{A comprehensive simulation framework for multi-modal kilonova observations from all-sky surveys}
\author[0000-0001-7129-1325]{Felipe Fontinele Nunes}
\affiliation{School of Physics and Astronomy, University of Minnesota, Minneapolis, Minnesota 55455, USA}
\affiliation{Theoretical Division, Los Alamos National Laboratory, Los Alamos, NM 87545, USA}
\email{}

\author[0009-0008-9546-2035]{Andrew Toivonen}
\affiliation{Theoretical Division, Los Alamos National Laboratory, Los Alamos, NM 87545, USA}
\email{atoivonen@lanl.gov}

\author[orcid=0000-0003-3481-5913]{Farhana Taiyebah}
\affiliation{Theoretical Division, Los Alamos National Laboratory, Los Alamos, NM 87545, USA}
\affiliation{Department of Scientific Computing, Florida State University, Tallahassee, FL 32306, USA}
\email{}

\author[orcid=0000-0002-5613-3019]{Leonard Lupin-Jimenez}
\affiliation{Theoretical Division, Los Alamos National Laboratory, Los Alamos, NM 87545, USA}
\affiliation{Department of Applied Mathematics, University of California, Santa Cruz}
\email{llupinji@ucsc.edu}

\author[orcid=0000-0003-3481-5913]{Skylar Callis}
\affiliation{Theoretical Division, Los Alamos National Laboratory, Los Alamos, NM 87545, USA}
\affiliation{Department of Mathematical Sciences, University of Delaware}
\email{sjcallis@udel.edu}

\author[orcid=0009-0007-4497-4674]{Malhar Kulkarni}
\affiliation{Department of Astrophysics, University of Virginia}
\email{jhz7az@virginia.edu}

\author[0000-0002-3316-5149]{Soumi De}
\affiliation{Theoretical Division, Los Alamos National Laboratory, Los Alamos, NM 87545, USA}
\affiliation{Center for Theoretical Astrophysics, Los Alamos National Laboratory, Los Alamos, NM 87545, USA}
\email{soumide@lanl.gov}

\author[0000-0002-8262-2924]{Michael W. Coughlin}
\affiliation{School of Physics and Astronomy, University of Minnesota, Minneapolis, Minnesota 55455, USA}
\email{cough052@umn.edu}

%\collaboration{all}{(PASP Editorial Office)}

\begin{abstract}
Current all-sky surveys such as Vera Rubin’s Legacy Survey of Space and Time and the Zwicky Transient Facility (ZTF) promise a wealth of scientific gains that are contained in the millions of astrophysical transient candidates produced each night. Kilonovae, one such transient of interest, will be challenging to identify in the alert stream and will require efficient artificial intelligence models to parse the large, real-time influx of data. In order to build these large models, comprehensive multimodal datasets are necessary for training. Due to a lack of numerous kilonova observations, we propose \texttt{kilonova-multimodal-emulator} -- a simulation pipeline for realistic, multimodal kilonova observations comprised of photometry, spectra, and images. We demonstrate this pipeline for ZTF-type observations, based on historical cadence and limiting magnitude information from the Bright Transient Survey (BTS) and using the latest radiative transfer kilonova models to produce a comprehensive dataset meant for the training of large artificial intelligence model.
\end{abstract}

%% PASP uses Unified Astronomy Thesaurus (UAT) concepts:
%\keywords{\uat{Galaxies}{573} --- \uat{Cosmology}{343} --- \uat{High Energy astrophysics}{739} --- \uat{Interstellar medium}{847} --- \uat{Stellar astronomy}{1583} --- \uat{Solar physics}{1476}}



\section{Introduction}

The multi-messenger detection of the kilonova AT2017gfo \citep{CoFo2017,Smartt:2017fuw,AbEA2017f} and GRB GRB170817A \cite{LIGOScientific:2017zic, Goldstein:2017mmi, Savchenko:2017ffs}, alongside \acp{GW} from the \ac{BNS} merger GW170817 \citep{AbEA2017b} revealed these events as sites of $r$-process nucleosynthesis via observation of heavy elements that power the kilonova emission \citep{LaSc1974,LiPa1998,MeMa2010,KaMe2017}. This discovery has led to great growth in the field of multi-messenger astronomy and scientific interest in additional observations. Kilonovae are short-lived astrophysical transients which may accompany neutron star mergers, specifically \ac{BNS} or \ac{NSBH} events. The joint detection of GW170817, AT2017gfo, and GRB170817A led to advances in nuclear astrophysics \citep{Margutti:2017cjl, Smartt:2017fuw, KaNa2017, KaMe2017,  Watson:2019xjv}, cosmology \citep{Abbott:2017xzu, Hotokezaka:2018dfi, Coughlin:2019vtv, Dietrich_2020, Bulla:2022ppy}, and tests of general relativity \citep{Ezquiaga:2017ekz, Baker:2017hug, Creminelli:2017sry}. Future observations promise additional advancements.

Kilonovae make challenging targets for observations, as they are short-lived and may be relatively faint. In addition, \ac{GW} sky localizations for promising neutron star mergers can span large sky localizations, specifically on the order of $\approx100-10,000 \ \mathrm{deg^2}$ \citep{Rover2007a, Fair2009,Fair2011,Grover:2013,WeCh2010,SiAy2014,SiPr2014,BeMa2015,EsVi2015,CoLi2015,KlVe2016}. %In order to maximize our chances of detection, we must use every means at our disposal. 
Alongside gravitational-wave follow-up, it is possible to identify kilonovae associated with sGRBs \citep{Tanvir:2013pia, AsCo2018, Jin:2019uqr}, long GRBs \citep{Rastinejad:2022zbg, Troja:2022yya, JWST:2023jqa}, or serendipitously through survey operations \citep{Andreoni:2021ykx, Almualla:2020ybm, Andrade:2025gme}. In this work, we will focus on serendipitous survey observations.

% peak of emissions within a day or two? cite GW170817
%https://arxiv.org/pdf/1710.05841 

%locating and identifying a transient as soon as possible is ideal in the hope of observing the peak of emissions. %There are numerous collaborations and telescopes currently devoting time and person-power towards finding the next \ac{EM} counterpart to a \ac{GW} event, and many more planned in the future.

There are numerous current and future telescopes that carry out wide-field surveys and participate in multi-messenger searches, such as the Panoramic Survey Telescope and Rapid Response System (Pan-STARRS) \citep{MoKa2012}, Asteroid Terrestrial-impact Last Alert System (ATLAS) \citep{ToDe2018}, the Zwicky Transient Facility (ZTF) \citep{Bellm2018,Graham2018,Ahumada:2024qpr}, BlackGEM \citep{BlGr2015}, Vera Rubin Observatory \citep{LSST:2008ijt}, the Nancy Grace Roman Space Telescope \citep{Andreoni:2023xlv}, ULTRASAT \citep{Shvartzvald:2023ofi}, and Ultraviolet Explorer (UVEX) \citep{Kulkarni:2021tit}.

In the absence of abundant observational data for kilonovae, models are necessary to simulate numerous kilonova observations across the parameter space. To generate photometry, or light curves, we employ a pair of models: i) \texttt{POSSIS}, the time-dependent, three-dimensional Monte Carlo radiative transfer code \citep{Bul2019} and ii) the \ac{LANL} TP2 model, which is a two-component, axisymmetric kilonova model consisting of toroidal dynamical ejecta and a peanut-shaped wind component \citep{Wollaeger:2021qgf}. These models are used to produce a grid of simulated kilonova light curves across the parameter space, from which a machine learning surrogate model is trained. This model allows for the simulation of models for arbitrary parameter values, as well as speeds up the calculation of the light curves and reduces the computational cost of simulation. Images and spectroscopic data will be generated to accompany this photometric data, with images simulated using realistic \ac{ZTF} \acp{PSF}.

There are existing efforts to create time-domain astronomy datasets for training AI models, such as the PLAsTiCC \citep{2018arXiv181000001T, Kessler:2019qge} and ELAsTiCC \citep{2023AAS...24111701N} time-series classification challenges. The PLAsTiCC and ELAsTiCC datasets contain a wide variety of time-domain signals, including supernovae, kilonovae, TDEs, active galactic nuclei (AGN), and many more. Designed as classification challenges, these datasets simulate realistic Vera C. Rubin Legacy of Space and Time Survey (LSST) photometric observations of the many transients of interest. In comparison, our dataset will be multi-modal, composed of spectra, photometry, and images. In this paper we will focus on ZTF observations encompassing all three modalities, but we have future plans to include LSST observations. The other key difference is we generate observations of kilonovae only for the scope of this paper; but those models are easily replaceable with the models for your source class of choice. Our simulations are intended not for a classification challenge to separate diverse classes of astrophysical sources, but instead for training an analysis architecture capable of identification and parameter estimation of a specific source class: in this case, kilonovae. 
%\rednote{do we want to compare to PLASTICC? https://arxiv.org/pdf/1810.00001, comparison to others works, ELASTIC, Applecider discusses this https://arxiv.org/pdf/2507.16088, make sure to cite, our dataset is multi-modal: correlated photometry, images, and spectra, why we are doing something different than plactic/elastic and why our work is novel}

With the LSST all-sky survey, Vera C. Rubin Observatory \citep{LSST:2008ijt} is ushering in a new era of multi-messenger astronomy: one that will see millions of alerts a night. AI and data science tools will be imperative to sort through the overwhelming number of transients to identify rare science-rich targets such as kilonovae. These AI tools will need to be trained on comprehensive and robust datasets, which this work will provide.

We devise a framework for the simulation of realistic observations from all-sky surveys, such as those carried out by the \ac{ZTF}. %and the Vera C. Rubin Observatory.
Specifically, we will simulate kilonova observations, but this pipeline can be used for generating mock datasets for other transient types as well by substituting the underlying models. Due to an abundance of real archival observations, we will focus on simulating \ac{ZTF} data, but with informed cadence, limiting magnitude, and other survey specific distributions, analogous simulations of kilonovae that would be observed by Vera C. Rubin could be carried out. For a comprehensive set of possible kilonova parameters, we will generate spectra, light curves, and images for our simulated kilonovae. The rest of this section demonstrates how each of those three observations are simulated. 

\section{DATA GENERATION: SPECTRA}
\label{sec: spectrum}
Spectroscopy carries information that the light curve alone does not. The time-dependent continuum shape, the strong color redistribution caused by line blanketing, and the wavelength dependence of host-galaxy contamination are all easier to represent in the spectral domain than through broadband fluxes alone. We generate a full observer-frame spectral sequence from the same kilonova SED family used below for the photometric simulations, evaluate that sequence at a sky position drawn from the host-associated galaxy pool used later for the image simulations, and, when desired, pass the result through a low-resolution instrument model. The spectroscopic layer is built in three stages: projection of the intrinsic kilonova SED to the observer frame, addition of the local host-galaxy spectrum at the selected position, and forward modeling through a spectrograph response.

\subsection{Kilonova spectral sequence}

We adopt the LANL spectral models presented by \citet{Wollaeger:2021qgf}, computed with the Monte Carlo radiative-transfer code \texttt{SuperNu} \citep{Wollaeger2013SuperNu,Wollaeger2014SuperNu}. The LANLTP2 library describes two-component, axisymmetric ejecta consisting of a lanthanide-rich toroidal dynamical component and a peanut-shaped wind with higher electron fraction, \(Y_e=0.27\). It samples the dynamical and wind masses, their characteristic expansion velocities, and the viewing direction. The same radiative-transfer family underlies the photometric simulations in the following section. The spectra and light curves therefore describe the same intrinsic emission at different levels of spectral resolution.

\subsubsection{Spectral emulator}
\label{subsec:spectral_emulator}

The LANLTP2 simulations sample a finite set of locations in a continuous physical parameter space. Repeating the Monte Carlo transport calculation for every synthetic event would make the construction of a large machine-learning dataset impractical. Fast surrogates for expensive spectral calculations have proved useful in several areas of astrophysics. In stellar and galaxy spectroscopy, the \textsc{Starfish} and \textsc{Speculator} frameworks compress synthetic spectral libraries into reduced representations and interpolate their coefficients with Gaussian processes or neural networks \citep{Czekala2015Starfish,Alsing2020Speculator}. Neural networks have also reproduced spectra from the \texttt{TARDIS} supernova radiative-transfer code with percent-level accuracy and a reduction of several orders of magnitude in evaluation time \citep{Kerzendorf2021Dalek}.

Kilonova-specific studies have used conditional variational autoencoders to emulate radiative-transfer spectra \citep{Lukosiute2022KilonovaNet}, Gaussian-process and adaptive-learning methods for light-curve interpolation \citep{Ristic2022}, and random forests to interpolate the time-dependent LANL spectra \citep{Ristic2023Spectra}. Spectra generated with a kilonova surrogate have subsequently supported amortized parameter inference for AT2017gfo on timescales of seconds \citep{Darc2024SpectralSBI}. These studies establish the practical value of treating the emergent spectrum as a structured function of the ejecta parameters. Our emulator combines a coordinate-conditioned nonlinear predictor, a wavelength-domain neural operator, and a reduced-order correction of the remaining errors. This construction retains the correlated evolution across wavelength, time, and viewing direction while providing the throughput required for population-scale data generation. A query to the emulator is specified by
\begin{equation}
\begin{aligned}
\mathbf{x}=\biggl(&
\log_{10}\frac{M_{\rm d}}{M_\odot},
\frac{v_{\rm d}}{c},
\log_{10}\frac{M_{\rm w}}{M_\odot},
\frac{v_{\rm w}}{c},\\
&\log_{10}\frac{t_{\rm rf}}{{\rm day}},
\mu\biggr),
\qquad
\mu\equiv |\cos\iota|.
\end{aligned}
\label{eq:spectral_emulator_input}
\end{equation}
where the subscripts \({\rm d}\) and \({\rm w}\) denote the dynamical and wind ejecta components.  Reflection symmetry about the equatorial plane allows the angular coordinate to be folded to \(0\leq\mu\leq1\).  The logarithmic coordinates for mass and time accommodate the several orders of magnitude covered by the diffusion and expansion timescales.  Before entering the neural network, each component of \(\mathbf{x}\) is standardized to a dimensionless interval centered on zero.  The training representation contains 66 rest-frame epochs from 0.125 to 34.90 days, 27 folded viewing directions, and 1024 logarithmically spaced wavelength bins from approximately 0.1 to 12.8~\(\mu\mathrm{m}\).

We represent each simulated directional spectrum by its bolometric luminosity and a normalized distribution of energy over the wavelength bins.  If \(f_{\lambda,j}\) is the flux density in wavelength bin \(j\), with bin width \(\Delta\lambda_j\), the spectral target is
\begin{equation}
p_j=
\frac{f_{\lambda,j}\Delta\lambda_j}
{\sum_k f_{\lambda,k}\Delta\lambda_k},
\qquad
p_j\geq0,
\qquad
\sum_j p_j=1.
\label{eq:emulator_spectral_probability}
\end{equation}
The network predicts a vector of logits \(\mathbf{z}\) and a standardized value of \(\ell_L=\log_{10}L_{\rm bol}\).  A softmax transformation, \(\widehat p_j=\exp z_j/\sum_k\exp z_k\), imposes positivity and unit normalization.  The flux density at the reference distance is reconstructed through
\begin{equation}
\widehat f_{\lambda,j,10\,\mathrm{pc}}
=
\frac{10^{\widehat\ell_L}}
{4\pi(10\,\mathrm{pc})^2}
\frac{\widehat p_j}{\Delta\lambda_j}.
\label{eq:emulator_flux_reconstruction}
\end{equation}
This representation gives the continuum shape and the overall luminosity separate numerical scales during optimization.  Equation~\eqref{eq:emulator_flux_reconstruction} restores their physical coupling at inference time and makes the wavelength integral of the reconstructed spectrum consistent with the predicted bolometric luminosity.

The first component of the emulator is a nonlinear predictor of \(\mathbf{z}\) and \(\ell_L\).  It combines a wavelength-local branch with a global latent branch.  Both branches use Fourier features of the normalized physical coordinates, which improve the representation of rapidly varying functions by low-dimensional coordinate networks \citep{Tancik2020FourierFeatures}.  For a scalar coordinate \(s\), the feature map has the form
\begin{equation}
\Phi_K(s)=
\left[
s,
\left\{\sin(\pi2^k s),\cos(\pi2^k s)\right\}_{k=0}^{K-1}
\right].
\label{eq:emulator_fourier_features}
\end{equation}
The local branch also embeds the logarithmic wavelength coordinate and propagates it through feature-wise linear modulation layers, following the general conditioning construction introduced by \citet{Perez2018FiLM}. If \(\mathbf{c}(\mathbf{x})\) is the encoded physical condition and \(\mathbf{h}_{j}^{(l)}\) is the hidden representation of wavelength bin \(j\), one layer applies
\begin{equation}
\mathbf{h}_{j}^{(l+1)}=
\sigma\!\left[
\mathbf{W}_l\mathbf{h}_{j}^{(l)}
\odot\left(1+0.1\boldsymbol{\gamma}_l(\mathbf{c})\right)
+0.1\boldsymbol{\beta}_l(\mathbf{c})
\right],
\label{eq:emulator_film}
\end{equation}
where \(\sigma\) is the SiLU activation and \(\boldsymbol{\gamma}_l\) and \(\boldsymbol{\beta}_l\) are learned functions of the ejecta parameters, time, and viewing direction.  This branch resolves wavelength-dependent changes in the broad opacity structure. The global branch adopts a latent-query construction related to \textsc{Perceiver IO} \citep{Jaegle2022PerceiverIO}. It converts the six physical inputs into parameter tokens, allows a set of learned latent vectors to attend to those tokens, and then uses the latent representation as the context for wavelength queries. It carries information between widely separated wavelength regions and represents the redistribution of energy across the full spectral energy distribution.  Learned scalar gates combine the local and global spectral logits and combine their luminosity predictions for every physical query.

A conditioned spectral operator refines this initial prediction.  The operator receives the standardized log probability and the logarithmic wavelength coordinate.  Six residual blocks combine a dilated depthwise convolution with a global Fourier convolution.  The Fourier contribution to one block can be written as
\begin{equation}
\mathcal{G}(\mathbf{h})
=
\mathcal{F}^{-1}
\left[
\mathbf{W}_k\,\mathcal{F}(\mathbf{h})_k
\right]_{k=0}^{K_{\rm op}-1},
\label{eq:emulator_spectral_operator}
\end{equation}
where \(\mathcal{F}\) denotes the discrete Fourier transform along wavelength, \(\mathbf{W}_k\) is a learned complex channel-mixing matrix, and the implementation retains \(K_{\rm op}=48\) modes. The dilated convolution responds to localized curvature and broad blended features. The Fourier term transmits corrections across the wavelength range. Feature-wise modulation by \(\mathbf{x}\) conditions every operator block on the physical state of the ejecta.

The learned Fourier convolution follows the neural-operator strategy of representing nonlocal interactions through a truncated set of spectral modes \citep{Li2021FNO,Kovachki2023NeuralOperator}. This approach has produced accurate function-valued surrogates for a broad class of parameterized physical systems. Comparative studies also show that the relative performance of operator architectures depends on the geometry, noise level, characteristic scales, and evaluation metric of the problem \citep{Lu2022NeuralOperators,Takamoto2022PDEBench}. We therefore retain the local dilated convolution together with the global Fourier pathway and assess the resulting spectra with separate luminosity, shape, multiscale, temporal, angular, and upper-tail diagnostics.

The neural predictor captures most of the spectral variation, and a reduced-order model is applied to its remaining errors. The combination of a proper orthogonal decomposition with a learned map from physical parameters to modal coefficients follows established non-intrusive reduced-order modeling methods \citep{Hesthaven2018PODNN}. Here the decomposition acts on the residual cube after the nonlinear prediction. Its modes consequently represent coherent error patterns that remain across wavelength, time, and viewing direction. For every training calculation \(n\), we assemble the complete residual cube
\begin{equation}
\begin{aligned}
R_n(t_a,\mu_b,\lambda_j)
&=\log\!\left[p_n(t_a,\mu_b,\lambda_j)+\epsilon\right]\\
&\quad-\log\!\left[p_n^{(0)}(t_a,\mu_b,\lambda_j)+\epsilon\right],
\end{aligned}
\label{eq:emulator_residual_cube}
\end{equation}
where \(p_n^{(0)}\) is the neural prediction and \(\epsilon=10^{-7}\) controls numerical behavior in negligible-flux bins.  The residuals are clipped at finite amplitude and centered over wavelength.  We flatten each time, angle, and wavelength cube into one vector and compute its proper orthogonal decomposition, which is a truncated singular-value decomposition of the collection of residual cubes.  Retaining \(R=160\) spectral modes gives
\begin{equation}
R_n(t,\mu,\lambda)
\simeq
\sum_{r=1}^{R}a_r(\boldsymbol{\vartheta}_n)
U_r(t,\mu,\lambda),
\label{eq:emulator_residual_decomposition}
\end{equation}
with
\begin{equation}
\boldsymbol{\vartheta}=
\left(
\log_{10}\frac{M_{\rm d}}{M_\odot},
\frac{v_{\rm d}}{c},
\log_{10}\frac{M_{\rm w}}{M_\odot},
\frac{v_{\rm w}}{c}
\right).
\end{equation}
Each mode \(U_r\) spans the entire wavelength, time, and viewing-angle domain.  The coefficient \(a_r\) belongs to one physical ejecta calculation and has no independent time or angular index.  This construction preserves the correlated temporal and angular evolution found by the radiative-transfer calculation.  At inference time, the stored modes are interpolated bilinearly in \(\log_{10}t\) and \(\mu\), after which the predicted coefficients reconstruct the residual spectrum at the requested epoch and line of sight.  A separate rank 48 decomposition applies the same construction to the standardized luminosity residual.

We obtain the residual coefficients from two complementary interpolants over \(\boldsymbol{\vartheta}\).  The neural interpolant first maps each normalized ejecta coordinate to Chebyshev polynomials.  The recurrence
\begin{equation}
\begin{aligned}
T_0(q)=1,
\qquad
T_1(q)=q,
\\
T_{m+1}(q)=2qT_m(q)-T_{m-1}(q)
\end{aligned}
\label{eq:emulator_chebyshev}
\end{equation}
is evaluated through degree six, and the resulting features are processed by a multilayer perceptron with separate heads for the spectral and luminosity coefficients.  Chebyshev features provide a well-conditioned polynomial description of smooth trends across the bounded mass and velocity domain.  The second interpolant is kernel regression.  For normalized ejecta coordinates \(\mathbf{q}\) and \(\mathbf{q}'\), we define the anisotropic distance
\begin{equation}
r^2(\mathbf{q},\mathbf{q}')
=
\sum_{m=1}^{4}
\frac{(q_m-q'_m)^2}{\ell_m^2}.
\end{equation}
The two residual models retained in the final emulator use Mat\'ern covariance functions
\begin{align}
k_{3/2}(r)&=\left(1+\sqrt{3}r\right)e^{-\sqrt{3}r},\\
k_{5/2}(r)&=\left(1+\sqrt{5}r+\frac{5r^2}{3}\right)e^{-\sqrt{5}r}.
\label{eq:emulator_matern}
\end{align}
For a matrix \(\mathbf{A}\) of known POD coefficients at the training nodes \(\mathbf{Q}\), the kernel prediction is
\begin{equation}
\widehat{\mathbf{a}}_{\rm ker}(\mathbf{q})
=
\mathbf{k}(\mathbf{q},\mathbf{Q})
\left[\mathbf{K}(\mathbf{Q},\mathbf{Q})+\eta\mathbf{I}\right]^{-1}
\mathbf{A}.
\label{eq:emulator_kernel_regression}
\end{equation}
The length scales \(\ell_m\), ridge parameter \(\eta\), and neural to kernel weight are selected using predictions in which the target physical simulation is excluded from the coefficient fit.  This cross-fitted procedure also estimates a multiplicative shrinkage for each mode.  Modes that reconstruct training residuals accurately but interpolate poorly receive smaller amplitudes.

The final model contains two independently fitted residual corrections, one using a Mat\'ern \(5/2\) covariance and one using a Mat\'ern \(3/2\) covariance.  Their normalized spectral predictions, \(p_j^{(1)}\) and \(p_j^{(2)}\), are combined through a smooth gate that depends on the physical query and wavelength,
\begin{equation}
b_j(\mathbf{x})
=
\operatorname{sigmoid}
\left[
\operatorname{logit}(b_0)
+s\tanh g_j(\mathbf{x})
\right],
\end{equation}
\begin{equation}
\widetilde p_j
=
\left[1-b_j(\mathbf{x})\right]p_j^{(1)}
+b_j(\mathbf{x})p_j^{(2)},
\qquad
\widehat p_j=\frac{\widetilde p_j}{\sum_k\widetilde p_k}.
\label{eq:emulator_probability_stack}
\end{equation}
The bounded correction around the initial mixture \(b_0=0.4\) limits rapid changes in the gate.  A smoothness penalty along wavelength further regularizes \(b_j\).  The luminosity output uses a fixed convex combination of the two luminosity predictions followed by a small conditioned correction.  The normalization in Equation~\eqref{eq:emulator_probability_stack} gives a positive spectral distribution for every query in the interpolation domain.

The loss function is defined in the same normalized energy coordinates used by the emulator.  This choice allows errors in spectral shape to be measured independently of errors in total luminosity, while additional terms couple the two quantities in physical flux.  Let \(n\) index the \(B\) spectra in a mini batch, let \(q_{nj}=\widehat p_{nj}\), and define the standardized luminosity target
\begin{equation}
y_n=\frac{\ell_{L,n}-\overline{\ell}_L}{s_L},
\qquad
\delta y_n=\widehat y_n-y_n,
\qquad
r_{L,n}=10^{s_L\delta y_n},
\label{eq:emulator_luminosity_residual}
\end{equation}
where \(s_L=0.91999\) dex is the standard deviation used to standardize \(\log_{10}L_{\rm bol}\).  Thus \(r_{L,n}=\widehat L_{{\rm bol},n}/L_{{\rm bol},n}\).  We use the smooth L1 function
\begin{equation}
\rho_\beta(u)=
\begin{cases}
u^2/(2\beta), & |u|\leq\beta,\\
|u|-\beta/2, & |u|>\beta,
\end{cases}
\label{eq:emulator_huber}
\end{equation}
which is quadratic near zero and linear for a large residual.  The main log spectral term is
\begin{equation}
\mathcal{L}_{\log p}
=
\frac{1}{B}\sum_{n=1}^{B}
\frac{
\sum_j \sqrt{p_{nj}^{\epsilon}}
\rho_{0.5}\!\left(\ln q_{nj}-\ln p_{nj}^{\epsilon}\right)
}{
\sum_j\sqrt{p_{nj}^{\epsilon}}
}.
\label{eq:emulator_log_shape_loss}
\end{equation}
Here \(p_{nj}^{\epsilon}=\max(p_{nj},10^{-12})\).  The softmax gives \(q_{nj}>0\), so only the simulated target requires a numerical floor.  Weighting by \(\sqrt{p_{nj}}\) retains sensitivity to the wings of the spectral energy distribution without allowing nearly empty wavelength bins to dominate the logarithmic residual.

Three complementary distances constrain the normalized energy distribution.  For an individual spectrum they are
\begin{align}
E_{{\rm shape},n}
&=\sum_j|q_{nj}-p_{nj}|,
\label{eq:emulator_shape_training}\\
H_n^2
&=\frac{1}{2}\sum_j
\left(\sqrt{q_{nj}}-\sqrt{p_{nj}}\right)^2,
\label{eq:emulator_hellinger_training}\\
E_{{\rm CDF},n}
&=\frac{1}{N_\lambda}\sum_j
\left|\sum_{k\leq j}(q_{nk}-p_{nk})\right|.
\label{eq:emulator_cdf_training}
\end{align}
The L1 distance \(E_{\rm shape}\) measures the total fraction of spectral energy assigned to the wrong wavelength bins.  The Hellinger distance gives a bounded comparison that remains stable when either distribution approaches zero.  The cumulative term measures the amount and distance of energy transport along the ordered wavelength axis.  A broad peak shifted coherently in wavelength therefore incurs a cumulative penalty even when its local residual is distributed over many bins.

The luminosity and spectral shape are joined through two flux losses.  The first operates on the logarithmic flux ratio,
\begin{equation}
\mathcal{L}_{\log F}
=\frac{1}{B}\sum_n
\frac{\sum_j\sqrt{p_{nj}^{\epsilon}}
\rho_{0.1}\!\left[
\log_{10}\!\left(\frac{q_{nj}}{p_{nj}^{\epsilon}}\right)
+s_L\delta y_n
\right]}
{\sum_j\sqrt{p_{nj}^{\epsilon}}},
\label{eq:emulator_log_flux_loss}
\end{equation}
and the second is the integrated physical flux error
\begin{equation}
E_{{\rm flux},n}
=\sum_j\left|r_{L,n}q_{nj}-p_{nj}\right|.
\label{eq:emulator_flux_training}
\end{equation}
Equation~\eqref{eq:emulator_flux_training} is also used as a validation statistic in Equation~\eqref{eq:emulator_validation_metrics}.  The central part of the training objective is
\begin{equation}
\begin{aligned}
\mathcal{L}_{\rm central}={}&
\mathcal{L}_{\log p}
+2.5\langle E_{\rm shape}\rangle
+0.5\langle H^2\rangle
+\langle E_{\rm CDF}\rangle\\
&+0.7\langle(\delta y)^2\rangle
+0.5\mathcal{L}_{\log F}
+2\langle E_{\rm flux}\rangle,
\end{aligned}
\label{eq:emulator_central_loss}
\end{equation}
where angle brackets denote an average over the mini batch.  The separate luminosity term fixes the normalization accurately, while the two joint terms prevent compensating errors in luminosity and normalized shape from producing an apparently favorable loss.

The broad features in LANLTP2 spectra span a range of wavelength scales.  We therefore aggregate adjacent wavelength bins in blocks of 4, 16, and 64 bins and compute the L1 distance between the predicted and target block probabilities at each resolution.  Their mean defines \(\mathcal{L}_{\rm ms}\).  We also form a six level, unnormalized Haar decomposition.  At each level, the detail coefficient is the difference between the probability in the left and right halves of a neighboring pair, while their sum is passed to the next level.  The mean L1 error in these detail coefficients defines \(\mathcal{L}_{\rm wav}\).  The first and second finite difference operators along logarithmic wavelength are denoted by \(D_1\) and \(D_2\).  We write their residuals as
\begin{equation}
\Delta_{r,nj}
=D_r\ln q_{nj}-D_r\ln p_{nj}^{\epsilon},
\qquad r\in\{1,2\}.
\end{equation}
The associated losses are
\begin{equation}
\mathcal{L}_{D_r}
=\frac{1}{B}\sum_n
\frac{\sum_j w_{nj}^{(r)}
\rho_{0.25}\!\left(\Delta_{r,nj}\right)}
{\sum_j w_{nj}^{(r)}}.
\label{eq:emulator_difference_loss}
\end{equation}
where \(w_{nj}^{(1)}=\sqrt{p_{n,j+1}+p_{nj}}\) and \(w_{nj}^{(2)}=\sqrt{p_{n,j+1}}\).  The resulting structural term is
\begin{equation}
\mathcal{L}_{\rm structure}
=0.25\mathcal{L}_{\rm ms}
+0.5\mathcal{L}_{\rm wav}
+0.1\mathcal{L}_{D_1}
+0.05\mathcal{L}_{D_2}.
\label{eq:emulator_structure_loss}
\end{equation}
The multiscale and wavelet terms constrain the energy contained in broad spectral regions.  The finite differences constrain the locations and widths of blended opacity features in log flux.

Each training item contains neighboring epochs and the complete set of folded viewing directions for one ejecta configuration.  We apply the weighted smooth L1 construction in Equation~\eqref{eq:emulator_difference_loss} to adjacent differences of \(\ln p\) along time and viewing direction, giving \(\mathcal{L}_{t,p}\) and \(\mathcal{L}_{\mu,p}\).  Smooth L1 losses on the corresponding adjacent differences of standardized log luminosity give \(\mathcal{L}_{t,L}\) and \(\mathcal{L}_{\mu,L}\).  Their contribution is
\begin{equation}
\mathcal{L}_{\rm sequence}
=0.2\left(\mathcal{L}_{t,p}+\mathcal{L}_{\mu,p}\right)
+0.1\left(\mathcal{L}_{t,L}+\mathcal{L}_{\mu,L}\right).
\label{eq:emulator_sequence_loss}
\end{equation}
This term constrains changes between neighboring spectra from the same transport calculation.  It is particularly useful during the rapid early color evolution and near lines of sight for which the projected column of lanthanide-rich material changes quickly.  At the midpoint between each pair of neighboring angular coordinates, we also evaluate the per spectrum loss against the arithmetic midpoint of the two target distributions and luminosities.  We denote this interpolation constraint by \(\mathcal{L}_{\rm mid}\).

A mean loss can give little leverage to a small population of difficult spectra. We therefore use an empirical upper-tail mean, closely related to conditional value at risk \citep{Rockafellar2000CVaR}, together with a smooth entropic-risk term. For an error vector \(e_1,\ldots,e_B\), these quantities are
\begin{equation}
\begin{aligned}
{\rm CVaR}_{0.075}(e)
&=\frac{1}{K}\sum_{n\in\mathcal{T}_K}e_n,
\qquad K=\lceil0.075B\rceil,\\
\mathcal{R}_{22}(e)
&=\frac{1}{22}\ln\!\left[
\frac{1}{B}\sum_{n=1}^{B}\exp(22e_n)
\right],
\end{aligned}
\label{eq:emulator_tail_risk}
\end{equation}
where \(\mathcal{T}_K\) contains the \(K\) largest errors in the mini batch.  The tail contribution used in training is
\begin{equation}
\begin{aligned}
\mathcal{L}_{\rm tail}={}&
{\rm CVaR}_{0.075}(E_{\rm shape})
+{\rm CVaR}_{0.075}(E_{\rm flux})\\
&+0.2\,{\rm CVaR}_{0.075}(|\delta y|)
+0.2\,\mathcal{R}_{22}(E_{\rm shape}).
\end{aligned}
\label{eq:emulator_tail_loss}
\end{equation}
The empirical tail means act directly on the least accurate spectra in each mini batch.  The differentiable entropic term distributes additional weight continuously toward larger shape errors.  Early spectra with \(t\leq0.355\) d and models with \(v_{\rm w}\leq0.08c\) are also sampled more frequently, since these regimes show faster spectral evolution and larger reconstruction errors.  Each condition increases the unnormalized sampling weight by 60\%, and both increments are retained where the conditions overlap.

For the wavelength dependent mixture in Equation~\eqref{eq:emulator_probability_stack}, we define \(\mathcal{L}_{\rm gate}=\langle|b_{j+1}-b_j|\rangle\).  The complete objective is
\begin{equation}
\begin{aligned}
\mathcal{L}={}&
\mathcal{L}_{\rm central}
+\mathcal{L}_{\rm structure}
+\mathcal{L}_{\rm sequence}\\
&+\mathcal{L}_{\rm tail}
+0.05\mathcal{L}_{\rm gate}
+0.05\mathcal{L}_{\rm mid}.
\end{aligned}
\label{eq:emulator_training_loss}
\end{equation}
All coefficients in Equation~\eqref{eq:emulator_training_loss} and its constituent terms are the values used for the reported emulator.  The two residual models are first optimized separately with this objective.  During the final combination fit their parameters remain fixed, and optimization updates the spectral gate and the small conditioned luminosity correction.  This prevents the combination stage from degrading either component while it learns where each residual interpolation is more accurate.

The design of the emulator reflects the scale hierarchy of spectra from expanding kilonova ejecta.  The large velocity gradient Doppler broadens and overlaps the dense forest of bound-bound transitions, producing wide troughs, shoulders, and peaks in the emergent spectrum.  The local branch and dilated convolutions resolve the wavelength dependence of these features.  The multiscale Fourier operator and the global latent branch carry information between the optical and infrared portions of the spectrum, where radioactive heating, photon diffusion, line blanketing, and the evolving ionization state redistribute the escaping energy as the ejecta expand.

Viewing direction changes the relative columns of the lanthanide-rich dynamical ejecta and the higher electron fraction wind along the line of sight.  It consequently changes the luminosity and color together, with the strongest angular variation occurring where the projected obscuration changes rapidly.  The complete residual cubes retain these correlations across wavelength, time, and angle within every POD mode.  Across the four ejecta parameters, the Chebyshev network supplies a stable global approximation over the bounded fitting domain.  The Mat\'ern \(5/2\) interpolant represents smoother coefficient variation, while the Mat\'ern \(3/2\) interpolant permits more rapid local changes.  Their conditioned mixture allows the preferred degree of smoothness to vary with physical state and wavelength.  Finally, the normalized probability representation enforces positivity and bolometric consistency, and the flux terms in the objective reconnect that representation to the physical luminosity scale used in the observer-frame projection.

\subsubsection{Projection to the observer frame}

Let
\begin{equation}
f^{\rm KN}_{\lambda,10\,{\rm pc}}
\left(t_{\rm rf},\lambda_{\rm rf};\boldsymbol{\vartheta},\iota\right)
\end{equation}
denote the rest-frame kilonova spectral flux density at a reference distance of \(10~{\rm pc}\), for intrinsic model parameters \(\boldsymbol{\vartheta}\) and viewing angle \(\iota\). Here \(\boldsymbol{\vartheta}\) collects the ejecta quantities that index the radiative-transfer model, while \(\iota\) fixes the line of sight through the axisymmetric ejecta. The \(10~{\rm pc}\) normalization isolates the spectral shape and temporal evolution from the later choice of source distance.

Let \(T_0\) be the merger time and \(\tau \equiv T-T_0\) the observer-frame time since merger. For an event observed at redshift \(z\), the rest-frame and observer-frame variables are related by
\begin{equation}
\lambda_{\rm obs}=(1+z)\lambda_{\rm rf},
\qquad
t_{\rm rf}=\frac{\tau}{1+z},
\end{equation}
so the observer samples a redshifted and time-dilated version of the same intrinsic sequence. The observer-frame spectrum is therefore
\begin{equation}
f^{\rm KN}_{\lambda,{\rm obs}}(\tau,\lambda_{\rm obs})
=
\frac{f^{\rm KN}_{\lambda,10\,{\rm pc}}}{1+z}
\left(
\frac{\tau}{1+z},
\frac{\lambda_{\rm obs}}{1+z};
\boldsymbol{\vartheta},\iota
\right)
\left(\frac{10~{\rm pc}}{D_L}\right)^2
,
\label{eq:spec_cosmo_projection}
\end{equation}
with \(D_L\) the adopted luminosity distance. The inverse-square term gives the geometric dimming. The extra \((1+z)^{-1}\) enters because \(f_\lambda\) is defined per unit wavelength, so stretching the wavelength interval by \((1+z)\) reduces the observed flux density by the same factor.

\subsubsection{Host-galaxy contamination}

The extracted spectrum includes light from the kilonova and its host. Once a host galaxy and an offset within that host have been selected, the transient is assigned a definite sky position \((\alpha_\star,\delta_\star)\), and the local galaxy spectrum must be estimated at that same position. The contaminating spectrum varies between the center of a dusty star-forming host and a faint outer arm or tidal feature. We use archival Pan-STARRS1 \(grizy\) imaging \citep{Chambers2016} and reconstruct the local galaxy spectrum with the chromatic scene-modeling approach introduced by HyperGal \citep{Lezmy2022}, using \texttt{CIGALE} as the galaxy SED-fitting engine \citep{Boquien2019}.

Around \((\alpha_\star,\delta_\star)\) we define a set of local spatial elements \(p\), each carrying a five-band photometric measurement \(F_{p,b}\) and its uncertainty for \(b\in\{g,r,i,z,y\}\). 
The images are registered to a common astrometric grid, and the fluxes are measured on a common spatial footprint so that the five values describe the same part of the galaxy. For a trial \texttt{CIGALE} spectrum \(S^{(m)}_p(\lambda)\), synthetic photometry in band \(b\) is obtained from the Pan-STARRS1 response \(T_b(\lambda)\). In a photon-counting convention this model flux is proportional to
\begin{equation}
F^{(m)}_{p,b}
=
\frac{\int S^{(m)}_p(\lambda)T_b(\lambda)\lambda\,d\lambda}
{\int T_b(\lambda)\lambda\,d\lambda}.
\label{eq:host_synthetic_photometry}
\end{equation}
The normalization and the physical SED parameters are constrained jointly by the five measured fluxes. The model library includes stellar-population evolution, star-formation history, nebular emission, and attenuation by dust. Each model receives a likelihood proportional to \(\exp(-\chi_m^2/2)\), with
\begin{equation}
\chi_m^2=\sum_b
\frac{\left(F_{p,b}-F^{(m)}_{p,b}\right)^2}{\sigma_{p,b}^2}.
\label{eq:host_cigale_likelihood}
\end{equation}
This construction uses \texttt{CIGALE} as a physically constrained interpolation from five broad bands to a continuous spectrum. Degeneracies among detailed stellar populations remain when only five-band photometry is available, and the adopted SED library sets the range of local spectra supported by the fit.

The result is a best-fitting observed-frame galaxy spectrum \(S_p(\lambda)\) for each retained spatial element. In practice, only spatial elements with sufficiently constrained multiband photometry are assigned independent SED fits. This avoids forcing arbitrary colors onto low-signal outskirts where the data do not support a meaningful local spectral reconstruction.

The contaminating host spectrum entering the transient extraction is then obtained by combining those local spectra with a kernel centered on the transient position,
\begin{equation}
H(\lambda)=\sum_p w_p\,S_p(\lambda).
\label{eq:host_local_spec}
\end{equation}
The weights \(w_p\) describe how much galaxy light from neighboring spatial elements contributes to the extracted transient spectrum. For seeing-limited follow-up, a natural choice is a Gaussian kernel,
\begin{equation}
w_p=\exp\!\left(-\frac{r_p^2}{2\sigma_{\rm see}^2}\right),
\qquad
\sigma_{\rm see}=\frac{{\rm FWHM}}{2\sqrt{2\ln2}},
\label{eq:host_spec_weights}
\end{equation}
where \(r_p\) is the projected distance between the transient position and the center of spatial element \(p\). 

The extraction weights also include the spatial-element area and the normalization appropriate to the adopted aperture. They therefore preserve the amount of host light entering the transient extraction, and their sum follows the aperture normalization. The same spatial kernel is applied at every wavelength before any instrument response is imposed. Spatial changes in color and line strength are retained through the individual \(S_p(\lambda)\), while the seeing kernel determines how those spectra mix at the selected position. The local fit includes regions with adequate multiband signal, for which the photometry supports a meaningful local spectral reconstruction.

Since \(H(\lambda)\) is reconstructed directly from observed Pan-STARRS1 photometry, it is already an observer-frame spectrum.

The high-resolution host-contaminated sequence generated by the emulator is therefore
\begin{equation}
f^{\rm tot}_{\lambda}(\tau,\lambda)
=
f^{\rm KN}_{\lambda,{\rm obs}}(\tau,\lambda)
+
H(\lambda).
\label{eq:host_contam_total_spec}
\end{equation}
The host term is fixed in wavelength for a given sky position, while the kilonova continuum changes rapidly with time. A useful diagnostic is the host fraction,
\begin{equation}
\phi_H(\tau,\lambda)
=
\frac{|H(\lambda)|}
{|f^{\rm tot}_{\lambda}(\tau,\lambda)|},
\label{eq:host_fraction_spec}
\end{equation}
which makes explicit that host contamination is both chromatic and phase dependent. Early in the evolution the kilonova may dominate the optical continuum, while at later phases the same extraction can become host dominated across much of the observed wavelength range. Figure~\ref{fig:spectrum_sequence} shows this progression for an injection position in NGC\,772, a nearby spiral at \(D_L\simeq 37.1\,\mathrm{Mpc}\) drawn from the same host pool used for the image simulations, using a relatively low-inclination kilonova realization with \(\iota\simeq 19^\circ\). 

\subsubsection{Instrumental response and observational effects: SEDM example}
\label{subsec:sedm_reduced_model}

\begin{figure*}[t]
  \centering
  \includegraphics[width=0.67\textwidth]{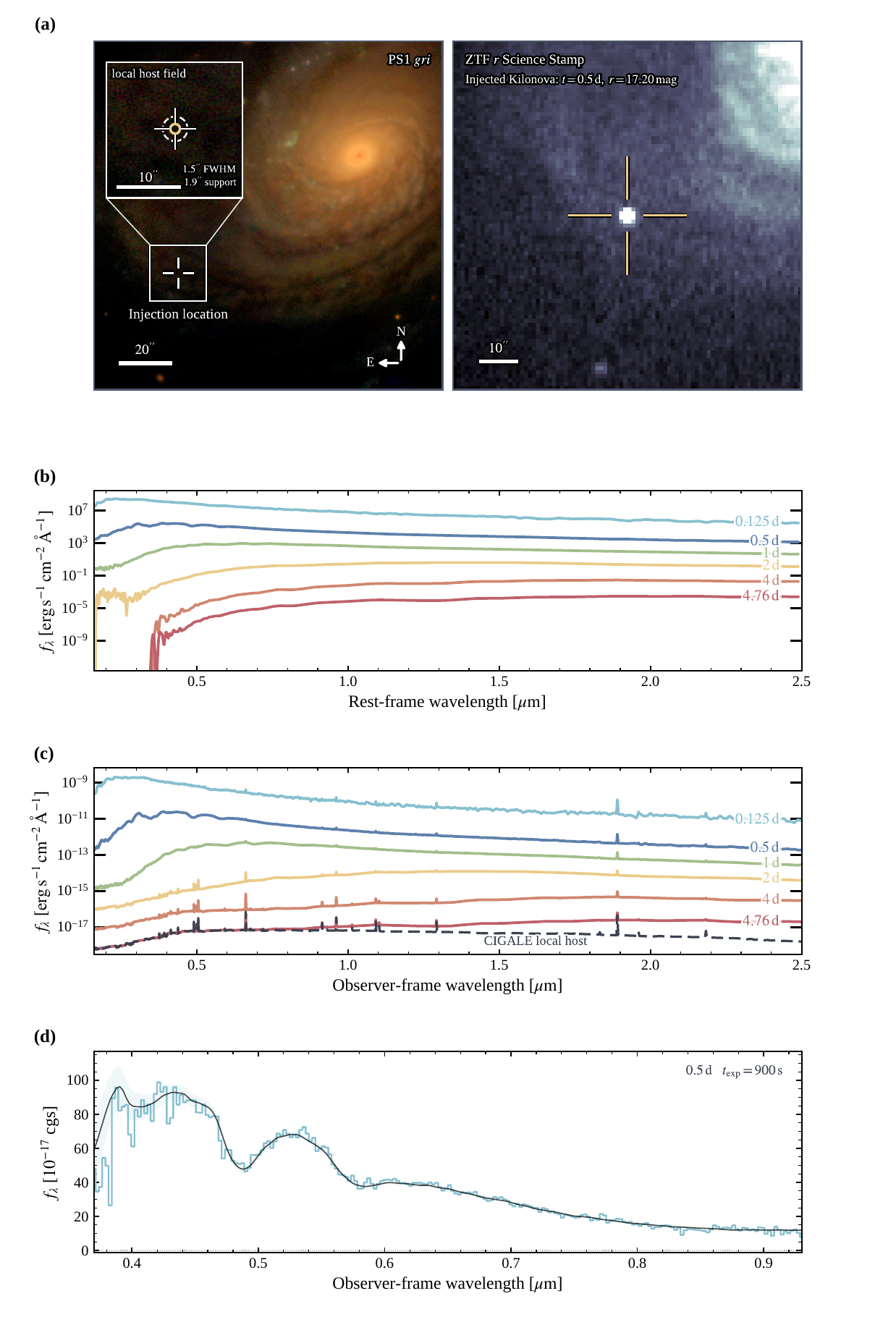}
  \caption{
    Example of the spectroscopic data-generation chain at an injection position in NGC\,772, at a projected offset of \(\simeq 42.9''\) from the galaxy center. Panel (a) shows the archival Pan-STARRS1 \(grizy\) cutouts centered on the selected sky position; these images provide the five-band measurements used by \texttt{CIGALE} to reconstruct the local host-galaxy spectrum entering Eq.~\eqref{eq:host_local_spec}. Panel (b) shows the transient sequence alone from the same \texttt{SuperNu} model family adopted for the photometric simulations, with viewing angle \(\iota\simeq 19^\circ\), plotted in the native rest-frame \(10~\mathrm{pc}\) \(f_\lambda\) normalization of the model grid. Panel (c) shows the corresponding observer-frame sequence after projection to the NGC\,772 distance \(D_L=37.1\,\mathrm{Mpc}\) and addition of the reconstructed local host spectrum, with both terms combined directly in the same \(f_\lambda\) units; the dashed black curve gives the host term by itself. The displayed phases run from \(0.125\) to \(4.76\) days after merger and are ordered from top to bottom from earlier to later times. The plotted spectra are displaced vertically by about \(0.7\) dex between successive epochs for clarity, with a slightly larger separation between the first two rest-frame phases. The local host continuum and narrow emission features are already evident during the first day, and by a few days after merger they dominate a large fraction of the optical-to-near-infrared wavelength range shown here.
  Panel (d) shows the flux-calibrated, host-subtracted SEDM-like realization of the \(0.5~\mathrm{d}\) source for a \(900~\mathrm{s}\) exposure at \(X=1.2\). The blue step curve is one noisy realization, the black curve is its noiseless reduced expectation, and the shaded band gives the reported \(1\sigma\) uncertainty.}
  \label{fig:spectrum_sequence}
\end{figure*}

The spectrum in Equation~\eqref{eq:host_contam_total_spec} retains the wavelength sampling of the kilonova model and the local host reconstruction. For panel~(d) of Figure~\ref{fig:spectrum_sequence}, we project this incident scene to the data level of a flux-calibrated, one-dimensional spectrum from the Spectral Energy Distribution Machine (SEDM). SEDM is a low-resolution integral-field spectrograph on the Palomar 60-inch telescope, designed for rapid transient classification at \(R\sim100\) \citep{Blagorodnova2018}. Its automated reduction produces calibrated spectra on a 220-bin optical wavelength grid \citep{Rigault2019}, while HyperGal models the point source and the spatially resolved host contribution within the SEDM cube \citep{Lezmy2022}.

We apply the instrumental line-spread function separately to the kilonova and host spectra and integrate the result over each finite SEDM wavelength bin. For a component \(q\in\{{\rm KN},H\}\),
\begin{equation}
\overline f^{\,q}_{k}
=
\sum_j M_{kj}(R)f^{\,q}_{\lambda,j},
\qquad
{\rm FWHM}(\lambda_j)=\frac{\lambda_j}{R},
\label{eq:sedm_finite_bin_projection}
\end{equation}
where \(j\) and \(k\) index the input and output bins, respectively, and \(M_{kj}\) is the response matrix obtained by integrating a normalized Gaussian line-spread function over both bins. We use \(R=100\) and a real SEDM wavelength grid covering approximately \(3700\) to \(9300~\text{\AA}\). The model spectrum has \(2~\text{\AA}\) sampling, which adequately samples the narrowest instrumental FWHM of \(37~\text{\AA}\). The reduced-spectrum mean and one stochastic realization are
\begin{equation}
\begin{aligned}
\mu_k &=
\left(\overline f^{\,{\rm KN}}_k+\eta_H\overline H_k\right)
\left[1+\delta_{{\rm cal},k}+\delta_{{\rm tel},k}\right],\\
\widehat f_k &= \mu_k+a\sigma_k\epsilon_k .
\end{aligned}
\label{eq:sedm_reduced_mean}
\end{equation}
Here \(\eta_H\) is the fraction of the projected host spectrum \(\overline H_k\) remaining after scene subtraction. The terms \(\delta_{{\rm cal},k}\) and \(\delta_{{\rm tel},k}\) describe smooth spectrophotometric residuals and residual O$_2$ and H$_2$O absorption, with the telluric templates evaluated using the \texttt{pysedm} airmass dependence. The reported uncertainty \(\sigma_k\), the valid-bin mask, the observation-wide scale \(a\), and the short-range correlation of the zero-centered, unit-variance Student-\(t\) deviate \(\epsilon_k\) are drawn from an empirical reduced-spectrum model. We calibrated that model using the pipeline-reported uncertainties and close-repeat differences from 2,988 nominal-quality SEDM spectra selected from the ZTF SN Ia DR2 and accessed through ZTFcosmo \citep{Rigault2025}. The fit uses uncertainty products and source-cancelled repeat differences, so it does not transfer Type Ia template residuals to the kilonova spectrum.

In our example, the projected \(0.5~\mathrm{d}\) kilonova has \(r=17.20\) AB mag. We adopt \(t_{\rm exp}=900~\mathrm{s}\), following the P60/SEDM exposure guide,\footnote{\url{https://sites.astro.caltech.edu/palomar/observer/60inchResources/p60sedm.html}} and evaluate the telluric residuals at \(X=1.2\). We set \(\eta_H=0\) to represent succesful subtraction of the HyperGal host model from the extracted mean. The incident host remains in the uncertainty calculation because its photons contribute to the local background. 

\section{DATA GENERATION: LIGHT CURVES}
%(...)

Photometric observations, commonly referred to as light curves when plotted over time, are an important tool for transient classification. The characteristics of kilonovae include a rise on the timescale of $\sim$day, where the light curve peaks and then fades quite rapidly, at a rate of $\sim$mag/day. This makes them fleeting targets for surveys and targeted follow-up alike. In order to generate realistic kilonova light curves observed by \ac{ZTF}, we will utilize informed distributions of limiting magnitudes and survey cadence, alongside physics-informed light curve models, to generate light curves as they would be seen by \ac{ZTF}. For the purposes of this paper we will focus on the \ac{LANL} TP2 model, a two-component 2D axisymmetric kilonova model comprised of a toroidal (T) dynamical
ejecta component and a spherical (S) wind component \citep{Wollaeger:2021qgf}, but one also can use the time-dependent, three-dimensional Monte Carlo radiative transfer code \texttt{POSSIS} \citep{Bul2019} for generating mock data with this pipeline.

%The final product is a finite set of discrete measurements and upper limits,
Our final product will be a finite set of discrete photometric observations and upper limits across three \ac{ZTF} bands:
\begin{equation}
\mathcal{D}
=
\left\{
\left(t_k,\; b_k,\; m_k,\; \sigma_k,\; m_{\lim,k}\right)
\right\}_{k=1}^{K},
\label{eq:lc_dataset_object}
\end{equation}
where \(t_k\) is the observation time (MJD), \(b_k\in\{g,r,i\}\) is the band, \(m_k\) is the reported magnitude (equal to the limiting magnitude for nondetections), \(\sigma_k\) is the reported magnitude uncertainty (infinite or a sentinel value for nondetections), and \(m_{\lim,k}\) is the single-epoch limiting magnitude.

\subsection{Radiative-transfer model grid and low-rank surrogate}
\label{subsec:svd_surrogate}

We generate intrinsic light curves from the LANLTP2 radiative-transfer model family \citep{Wollaeger:2021qgf}. These axisymmetric models describe a toroidal dynamical-ejecta component and a peanut-shaped wind component, with the wind composition fixed to the model's \(Y_e=0.27\) prescription. For the photometric surrogate, we define the intrinsic parameter vector \(\boldsymbol{\vartheta}\in\Theta\subset\mathbb{R}^{5}\) as
\begin{equation}
\boldsymbol{\vartheta}
=
\left(
\log_{10}\frac{M_{\rm d}}{M_\odot},
\frac{v_{\rm d}}{c},
\log_{10}\frac{M_{\rm w}}{M_\odot},
\frac{v_{\rm w}}{c},
\iota
\right),
\end{equation}
where \(M_{\rm d}\) and \(v_{\rm d}\) are the mass and characteristic expansion velocity of the dynamical ejecta, \(M_{\rm w}\) and \(v_{\rm w}\) describe the wind, and \(\iota\) is the viewing angle measured from the symmetry axis. The ejecta morphology and composition remain fixed within this model family; the surrogate interpolates across the masses, velocities, and viewing directions represented by the simulations.

The radiative-transfer library provides rest-frame absolute AB magnitudes in each band \(b\in\mathcal{B}\), evaluated at discrete parameter configurations \(\{\boldsymbol{\vartheta}^{(n)}\}_{n=1}^{N_{\mathrm{grid}}}\) and a fixed rest-frame time grid \(\{t_j\}_{j=1}^{J}\subset[t_{\min},t_{\max}]\):
\begin{equation}
M_b^{(n)}(t_j)
\equiv
M_b(t_j;\boldsymbol{\vartheta}^{(n)}).
\end{equation}

\begin{figure*}[t]
  \centering
  \includegraphics[width=\textwidth]{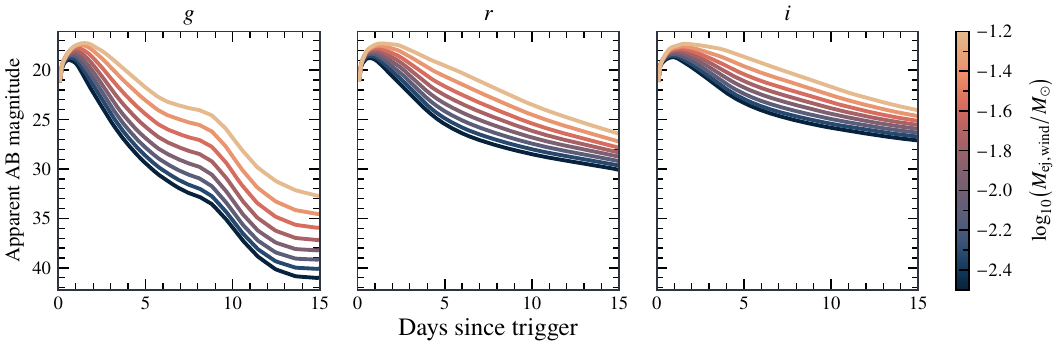}
  \caption{
    Example model light-curve families in ZTF $gri$ bands as a function of the mass of the wind ejecta $\log_{10} M_{\rm ej,wind}$, an intrinsic parameter. Other intrinsic and extrinsic injection parameters are fixed and consistent with AT2017gfo as infered by the maximum likelihood estimate from \cite{King_2025} for the LANLTP2 model. Curves are shown as apparent AB magnitudes.}

  \label{fig:lc_families}
\end{figure*}

Radiative transfer simulations are prohibitive in time and computational expense when used for kilonova template generation within a parameter estimation pipeline. %Direct interpolation in \(\Theta\) at runtime is computationally expensive and scales poorly with \(N_{\mathrm{grid}}\) and \(J\). 
We therefore compress the model grid into a low-dimensional representation via a truncated singular-value decomposition (SVD), performed independently for each band~\citep{pang22}. This model can then quickly and cheaply simulate kilonova light curves from arbitrary parameters, allowing for the construction of large datasets suitable for training a machine learning architecture. Example simulated light curves from the surrogate model can be seen in Figure~\ref{fig:lc_families}.

For a fixed band \(b\), define the matrix \(\mathbf{Y}_b\in\mathbb{R}^{N_{\mathrm{grid}}\times J}\) with entries \((\mathbf{Y}_b)_{n j} = M_b(t_j;\boldsymbol{\vartheta}^{(n)})\). We apply an affine rescaling in time to improve conditioning,
\begin{equation}
\tilde{\mathbf{Y}}_b(:,j)
=
\frac{\mathbf{Y}_b(:,j) - \alpha_{b,j}}{\beta_{b,j}},
\qquad
\beta_{b,j}>0,
\label{eq:affine_time_norm}
\end{equation}
where \(\alpha_{b,j}\) and \(\beta_{b,j}\) are deterministic normalization constants computed from the grid (e.g., time-wise minima and ranges, or means and standard deviations). We then compute the SVD
\begin{equation}
\tilde{\mathbf{Y}}_b
=
\mathbf{U}_b\,\mathbf{S}_b\,\mathbf{V}_b^{\top},
\end{equation}
and retain the first \(K_b\) components, with \(K_b\ll J\). Writing \(\mathbf{V}_{b,K}\in\mathbb{R}^{J\times K_b}\) for the retained right-singular vectors and \(\mathbf{c}_{b}^{(n)}\in\mathbb{R}^{K_b}\) for the corresponding coefficient vector of the \(n\)-th grid point, we obtain the approximation
\begin{equation}
\tilde{M}_b(t_j;\boldsymbol{\vartheta}^{(n)})
\approx
\sum_{\ell=1}^{K_b} c_{b,\ell}^{(n)}\,\phi_{b,\ell}(t_j),
\qquad
\phi_{b,\ell}(t_j)\equiv (\mathbf{V}_{b,K})_{j\ell}.
\label{eq:svd_expansion_discrete}
\end{equation}
In continuous time, we interpret \(\phi_{b,\ell}(t)\) as basis functions obtained by interpolating \(\phi_{b,\ell}(t_j)\) over \(t\in[t_{\min},t_{\max}]\). The surrogate model then takes the form
\begin{equation}
M_b(t;\boldsymbol{\vartheta})
\approx
\alpha_b(t) + \beta_b(t)\,
\sum_{\ell=1}^{K_b} c_{b,\ell}(\boldsymbol{\vartheta})\,\phi_{b,\ell}(t),
\label{eq:svd_surrogate_continuous}
\end{equation}
where \(\alpha_b(t)\) and \(\beta_b(t)\) are the continuous-time counterparts of the normalization in Eq.~\eqref{eq:affine_time_norm}. The remaining task is to predict the coefficient map \(\boldsymbol{\vartheta}\mapsto \mathbf{c}_b(\boldsymbol{\vartheta})\) at arbitrary \(\boldsymbol{\vartheta}\) within the region spanned by the grid. We learn this map by training a regression model \(\mathcal{R}_b:\Theta\rightarrow\mathbb{R}^{K_b}\) on the paired set \(\{(\boldsymbol{\vartheta}^{(n)},\mathbf{c}_b^{(n)})\}\), yielding
\begin{equation}
\mathbf{c}_b(\boldsymbol{\vartheta}) \approx \mathcal{R}_b(\boldsymbol{\vartheta}).
\end{equation}
Equation~\eqref{eq:svd_surrogate_continuous} then defines a fast, differentiable approximation to the radiative-transfer light curve in each band. While our implementation also tracks bolometric luminosity using an analogous low-rank representation, the dataset used for the present work is defined by the band-limited magnitudes \(M_b(t;\boldsymbol{\vartheta})\) for \(b\in\{g,r,i\}\).

\subsection{Extrinsic parameters and observer-frame ``truth'' light curves}
\label{subsec:observer_truth}

Each synthetic event is further specified by a set of extrinsic (observer-dependent) parameters,
\begin{equation}
\boldsymbol{\xi} \equiv (d_L,\; z,\; E,\; T_0),
\end{equation}
where \(d_L\) is luminosity distance, \(z\) is redshift, \(E\equiv E(B-V)\) is color excess, and \(T_0\) is the trigger time (MJD) that anchors the event to the survey timeline. We adopt a physically motivated prior in which distances are sampled uniformly in proper volume within \([D_{\min},D_{\max}]\), 
\begin{equation}
p(d_L)
=
\frac{3 d_L^2}{D_{\max}^3-D_{\min}^3},
\qquad
d_L\in[D_{\min},D_{\max}],
\label{eq:distance_prior_paper}
\end{equation}
and the redshift is deterministically assigned by inverting the cosmological luminosity-distance relation \(d_L(z)\) under a fixed background cosmology, i.e.\ \(z=z(d_L)\). We draw \(E\) from a modest-support distribution and draw \(T_0\) uniformly over a survey time window \([T_{\min},T_{\max}]\), so that events occur at random phases with respect to the survey cadence.

Given \((\boldsymbol{\vartheta},\boldsymbol{\xi})\), we define a continuous-time observer-frame ``truth'' magnitude in each band. Let \(\tau\equiv T-T_0\) denote observer-frame time since trigger. Cosmological time dilation implies that the corresponding rest-frame time is \(\tau/(1+z)\). We therefore evaluate the surrogate in rest-frame time and then apply distance dimming, extinction, and a conventional \((1+z)\) photometric factor consistent with mapping rest-frame spectral-density conventions into observed AB magnitudes when detailed K-corrections are neglected at low redshift. Collecting these effects, we define
\begin{equation}
m_b^{\mathrm{true}}(\tau;\boldsymbol{\vartheta},\boldsymbol{\xi})
=
M_b\!\left(\frac{\tau}{1+z};\boldsymbol{\vartheta}\right)
+
\mu(d_L)
+
A_b(E,z)
+
\Delta_b(z),
\label{eq:true_mag_definition}
\end{equation}
where \(\mu(d_L)\) is the distance modulus,
\begin{equation}
\mu(d_L)=5\log_{10}\!\left(d_L[\mathrm{Mpc}]\right)+25,
\label{eq:distance_modulus}
\end{equation}
\(A_b(E,z)\) is the band-dependent dust attenuation in magnitudes computed from an extinction law evaluated at the appropriate effective wavelength (including the shift of the effective rest wavelength by \(1+z\)), and \(\Delta_b(z)\) denotes the adopted small redshift-dependent magnitude correction term. Therefore, equation~\eqref{eq:true_mag_definition} defines a smooth, noise-free apparent-magnitude curve on \(\tau\in[t_{\min},t_{\max}]\).

At fixed intrinsic and extrinsic parameters, the surrogate produces a deterministic light curve. We optionally apply data augmentation by adding small magnitude perturbations before sampling the curve at the survey epochs. On the dense observer-frame time grid \(\{\tau_j\}\), we define
\begin{equation}
m_{b,j}^{\mathrm{dense}}
=
m_b^{\mathrm{true}}(\tau_j;\boldsymbol{\vartheta},\boldsymbol{\xi})
+
\epsilon_{b,j},
\qquad
\epsilon_{b,j}\sim\mathcal{N}(0,\sigma_{\mathrm{model}}^2).
\label{eq:model_scatter}
\end{equation}
In this prescription, the perturbations are independent across bands and dense-grid epochs. They broaden the distribution of generated light curves around each surrogate prediction, allowing downstream models to encounter small departures from the exact templates during training.

The parameter \(\sigma_{\mathrm{model}}\) controls the augmentation amplitude and is separate from the photometric measurement uncertainty assigned later to each survey visit. The augmentation changes the generated magnitudes at fixed ejecta parameters; the physical parameter domain remains the one covered by the radiative-transfer library. Independent Gaussian perturbations provide a simple prescription for local variation around the templates, although physical changes in ejecta properties can produce correlated changes across time and wavelength.  Setting \(\sigma_{\mathrm{model}}=0\) recovers the unperturbed surrogate light curve.

\subsection{Data-driven survey sampling: cadence, depth, and band switching}
\label{subsec:survey_sampling_operator}

\begin{figure*}[t]
  \centering
  \includegraphics[width=\textwidth]{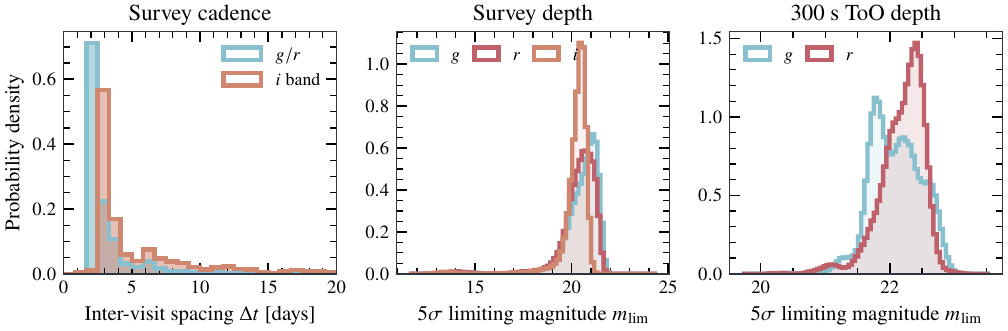}
    \caption{
    Survey-model components used to construct the synthetic ZTF observations.
    \textit{Left:} empirical inter-visit spacing distributions for the public \(g/r\) survey cadence and the separately modeled \(i\)-band revisit cadence.
    \textit{Middle:} single-epoch \(5\sigma\) limiting-magnitude distributions measured from the public-program observations, shown separately in \(g\), \(r\), and \(i\).
    \textit{Right:} single-epoch \(5\sigma\) limiting-magnitude distributions for 300\,s ToO observations in \(g\) and \(r\).
    }
    \label{fig:cadence_depth}
\end{figure*}

The mapping from a smooth light curve \(m_b^{\mathrm{dense}}(\tau)\) to a discrete dataset \(\mathcal{D}\) is a central step in producing survey-realistic simulated observations. We represent this mapping as a stochastic observation operator that generates a sequence of visits \(\{(t_k,b_k,m_{\lim,k})\}\), where \(t_k\) is the observer-frame epoch, \(b_k\) is the observed band, and \(m_{\lim,k}\) is the single-epoch limiting magnitude. Rather than imposing an idealized regular cadence or a fixed survey depth, the distributions entering this operator are inferred from historical ZTF observations. We follow the data-driven survey-modeling approach of \citet{Sravan2022}, with the ZTF Phase II implementation described by \citet{Kiendrebeogo:2023hzf}. In this construction, the routine public-survey cadence and the separately modeled \(i\)-band cadence are represented empirically, while the variation in single-epoch depth is likewise drawn from distributions measured from ZTF observations. The cadence and limiting-magnitude distributions used in our forward model are shown in Figure~\ref{fig:cadence_depth}.

We generate visit times directly in the observer frame. For each simulated event, the first observation occurs at a randomly sampled offset \(\tau_{\mathrm{start}}>0\) after merger. Subsequent epochs are generated by drawing revisit intervals from kernel-density estimates of the empirical ZTF cadence distributions. Following \citet{Sravan2022,Kiendrebeogo:2023hzf}, the routine public-survey cadence and the \(i\)-band revisit cadence are modeled separately. The observing filter is associated with the sampled visit sequence rather than drawn independently at every epoch, preserving the empirical relationship between visit timing and filter sequence present in the survey history. This procedure reproduces the broad and irregular distribution of ZTF sampling intervals, including closely separated observations and longer gaps between successive visits, without reducing the survey to a single characteristic cadence.

For simulations that include rapid follow-up, we augment the baseline survey sampling with a target-of-opportunity (ToO) component motivated by historical ZTF follow-up of gravitational-wave events. These observations are treated separately from the routine survey cadence and use the corresponding longer-exposure configurations. In particular, the survey model includes 300\,s \(g\)- and \(r\)-band observations representative of the ZTF gravitational-wave follow-up strategy described by \citet{Kiendrebeogo:2023hzf}. The resulting observation process can therefore combine routine survey sampling with a distinct early-time follow-up component while retaining the same underlying physical kilonova realization.

Once the visit times and bands have been generated, each observation is assigned a limiting magnitude \(m_{\lim,k}\) drawn from an empirical, band-dependent depth distribution. The observed ZTF distributions for cadence and limiting magnitude are shown in Figure~\ref{fig:cadence_depth}. We model this quantity stochastically rather than assigning a single characteristic ZTF depth because the sensitivity of individual exposures varies substantially with observing conditions. The distributions used here are derived from historical ZTF measurements and are constructed separately for the relevant survey components and filters. For the ToO component, the limiting-magnitude distribution is conditioned on the longer exposure configuration. These empirical distributions are shown in the middle and right panels of Figure~\ref{fig:cadence_depth}. The resulting survey realization is
\begin{equation}
\mathcal{S}(\boldsymbol{\xi})
=
\{(t_k,b_k,m_{\lim,k})\}_{k=1}^{K},
\end{equation}
with \(t_k\in[T_0+t_{\min},T_0+t_{\max}]\).

Given \(\mathcal{S}\), we evaluate the dense-time magnitudes by interpolation. Let \(\tau_k=t_k-T_0\). We assign the noiseless apparent magnitude at each visit as
\begin{equation}
m_k^{\star}
=
m_{b_k}^{\mathrm{dense}}(\tau_k),
\label{eq:visit_eval}
\end{equation}
where the right-hand side is obtained by interpolating the dense grid \(\{(\tau_j,m_{b,j}^{\mathrm{dense}})\}\).

\subsection{Photometric uncertainties, detections, and upper limits}
\label{subsec:noise_and_censoring}

% "We" would be the correct here
%\rednote{The not We??}

We model reported uncertainties using a magnitude-dependent, band-dependent distribution calibrated to survey photometry. Specifically, for each band \(b\) we fit a flexible parametric family to the conditional distribution \(p(\sigma_m\mid m,b)\). In practice we discretize magnitude into bins \([m_{\ell},m_{\ell+1})\) and, in each bin, represent the distribution of reported magnitude errors by a skew-normal (or similarly flexible) law with parameters that vary with \(b\) and bin index. We denote the resulting conditional law abstractly by
\begin{equation}
\sigma_k \sim p(\sigma_m \mid m_k^{\star}, b_k),
\label{eq:sigma_conditional}
\end{equation}
optionally augmented by an error floor added in quadrature to represent irreducible calibration systematics.

The detection model is then expressed in terms of the limiting magnitude \(m_{\lim,k}\). If the source is fainter than the limiting magnitude, we treat the visit as a nondetection and record it as upper limit by setting $\sigma_k$ to infinity:
\begin{equation}
\text{if } m_k^{\star} > m_{\lim,k}
\quad\Rightarrow\quad
m_k = m_{\lim,k},
\qquad
\sigma_k = \infty.
\label{eq:upper_limit_encoding}
\end{equation}
If the source is brighter than the limiting magnitude, we treat the visit as a detection and draw a noisy magnitude measurement. In magnitude space we adopt a Gaussian perturbation,
\begin{equation}
\text{if } m_k^{\star} \le m_{\lim,k}
\quad\Rightarrow\quad
m_k \sim \mathcal{N}(m_k^{\star},\sigma_k^2),
\label{eq:gaussian_mag_noise}
\end{equation}
with \(\sigma_k\) drawn from Eq.~\eqref{eq:sigma_conditional}. The use of a Gaussian in magnitudes is an approximation to flux-space noise that is accurate at moderate-to-high signal-to-noise; the survey-calibrated distribution for \(\sigma_k\) partially absorbs deviations from this idealization.

\begin{figure}[t]
  \centering
  \includegraphics[width=\columnwidth]{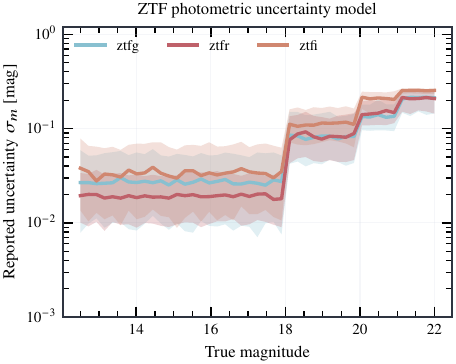}
  \caption{
  Empirical ZTF photometric uncertainty model: reported magnitude uncertainty $\sigma_m$ versus true magnitude for $gri$.
  The solid curves trace the median $\sigma_m$; shaded regions enclose the central 68\% probability mass of the conditional distribution in Eq.~\eqref{eq:sigma_conditional}.}
  \label{fig:ztf_uncertainty}
\end{figure}

Figure~\ref{fig:ztf_uncertainty} shows the empirical magnitude-dependent uncertainty model used in this step. The typical reported uncertainty increases toward fainter magnitudes, while the finite width of the conditional distributions retains the observed scatter in photometric precision at fixed source brightness. Modeling \(p(\sigma_m\mid m,b)\) in this way allows the simulation to reproduce the heteroscedastic character of ZTF photometry.

This construction yields a mixed dataset of detections and nondetections. If one wishes to write the implied likelihood for the entire light curve given \((\boldsymbol{\vartheta},\boldsymbol{\xi})\), it factorizes over visits into a product of Gaussian terms for detections and survival terms for censored observations. Denoting by \(\mathcal{I}_{\mathrm{det}}\) and \(\mathcal{I}_{\mathrm{lim}}\) the index sets of detections and limits, respectively, we may write schematically 

% Formatting fixed on this equation

\begin{equation}
\begin{split}
p(\mathcal{D}\mid \boldsymbol{\vartheta},\boldsymbol{\xi})
={}&
\prod_{k\in\mathcal{I}_{\mathrm{det}}}
\mathcal{N}\!\left(
m_k\,\big|\,m_{b_k}^{\mathrm{true}}(\tau_k),\sigma_k^2
\right)
\\
&\times
\prod_{k\in\mathcal{I}_{\mathrm{lim}}}
\mathbb{P}\!\left(
m>m_{\lim,k}\,\big|\,m_{b_k}^{\mathrm{true}}(\tau_k)
\right).
\end{split}
\label{eq:full_lc_likelihood}
\end{equation}

where the second factor represents censoring at the limiting magnitude. The stored representation in Eq.~\eqref{eq:lc_dataset_object} is thus a compact encoding of a censored heteroscedastic time series, aligned to the survey cadence.

\begin{figure*}[t]
  \centering
  \includegraphics[width=\textwidth]{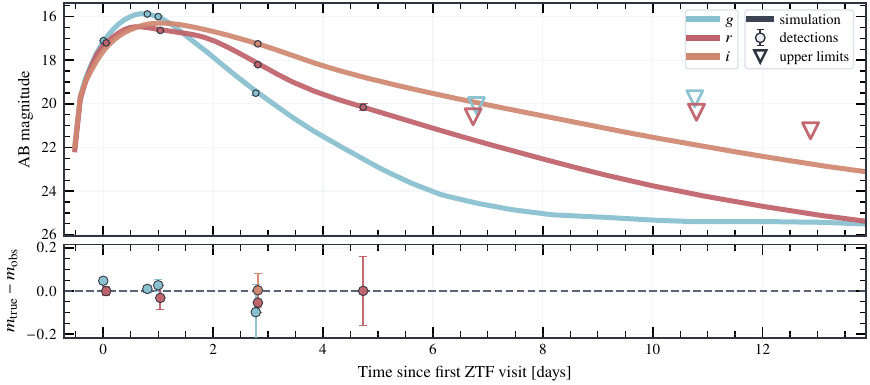}
  \caption{
    Example injected kilonova light curve (solid lines) and ZTF-like survey sampling (markers)
    in $gri$ bands. Upper limits are shown as open inverted triangles.
    Bottom panel shows residuals $m_{\rm true}-m_{\rm obs}$ for detections. Intrinsic and extrinsic injection parameters are consistent with AT2017gfo as infered by the maximum likelihood estimate from \cite{King_2025} for the LANLTP2 model.
    }
  \label{fig:truth_vs_ztf}
\end{figure*}

The combined effect of the survey forward model is illustrated in Figure~\ref{fig:truth_vs_ztf} using an AT2017gfo-like realization from the LANLTP2 model family. We adopt the maximum-likelihood parameters inferred by \citet{King_2025}, with a dynamical-ejecta mass \(M_{\rm d}=0.040,M_\odot\) and velocity \(v_{\rm d}=0.27c\), a wind-ejecta mass \(M_{\rm w}=0.028,M_\odot\) and velocity \(v_{\rm w}=0.07c\), and a viewing angle \(\iota=0.75\) rad. The central realization is placed at a luminosity distance \(D_L=44\) Mpc. The solid curves show the resulting continuous observer-frame \(gri\) light curves, while the survey forward model converts them into a sparse set of detections and upper limits with epochs, filters, limiting depths, and photometric uncertainties sampled from the empirical ZTF model. The lower panel shows the residuals \(m_{\rm true}-m_{\rm obs}\) for detected observations. Thus, even for a fixed underlying kilonova realization, different realizations of the observation operator produce different measured light curves, allowing the simulated dataset to capture the observational variability encountered by downstream classification and inference models.

\subsection{Dataset realization and interface to image-level injections}
\label{subsec:dataset_interface}

Repeating the procedure above for \(N_{\mathrm{inj}}\) independent draws \((\boldsymbol{\vartheta}_i,\boldsymbol{\xi}_i)\) yields a Monte Carlo dataset
\begin{equation}
\mathcal{T}
=
\left\{
\left(\boldsymbol{\vartheta}_i,\boldsymbol{\xi}_i,\mathcal{D}_i\right)
\right\}_{i=1}^{N_{\mathrm{inj}}}.
\end{equation}
Each element \(\mathcal{D}_i\) consists of a realistic mixture of detections and upper limits across multiple bands, with times and depths that mimic the empirical survey sampling.

\section{DATA GENERATION: IMAGES}
\label{sec:psf_injection}

We now turn to the image-level component of the simulation. The input to this stage is an unresolved transient at sky position \((\alpha_\star,\delta_\star)\) with an observer-frame apparent magnitude \(m_{e,\star}\) specified for one or more survey epochs \(e\). The goal is to generate realistic \ac{ZTF} image products for those epochs by inserting the transient directly into archival science images and pairing the result with the corresponding archival reference images. In this construction the transient is the only synthetic ingredient. The host-galaxy light, crowding, sky background, detector artifacts, and the full diversity of the survey imaging are all inherited from real \ac{ZTF} data products \citep{Bellm2018,Graham2018,Masci2019}. We see an example of the final product in Figure~\ref{fig:gw170817_sequence}.

We also retain the reference image as an explicit product rather than collapsing the problem immediately to image subtraction. At each epoch we therefore construct a science/reference pair,
\begin{equation}
\mathcal{I}_e \equiv \left(\mathbf{I}^{\rm sci}_{e,\mathrm{inj}},\widetilde{\mathbf{I}}^{\rm ref}_e\right),
\end{equation}
where \(e\) labels a specific archival exposure, \(\mathbf{I}^{\rm sci}_{e,\mathrm{inj}}\) is the science image after transient injection, and \(\widetilde{\mathbf{I}}^{\rm ref}_e\) is the corresponding archival reference image after alignment to the science-image frame. This operator is completely general. It can be used for any unresolved transient, independent of how the apparent magnitudes were generated. Kilonova specificity enters only through the astrophysical prior used to choose the host environment and source location, together with the externally supplied light curve \(m_b(t)\).

\subsection{Single-epoch science/reference stamps construction}
\label{subsec:image_single_epoch}

Consider an archival \ac{ZTF} science exposure \(e\), observed at time \(t_e\) in filter \(b_e\in\{g,r,i\}\), together with its archival reference image for the same field, detector quadrant, and filter. This is the same basic setting used in survey pipelines and in image-level transient simulations more generally, where synthetic point sources are overlaid on real survey images and interpreted relative to a static reference scene in order to assess detection and measurement performance \citep{Bellm2018,Graham2018,Masci2019,Kessler2015DiffImg,Goldstein2015AutoScan,Kessler_2009,Kessler:2019qge}. The reference image is first aligned to the science exposure and placed on the same flux scale:
\begin{equation}
\widetilde{\mathbf{I}}^{\rm ref}_e
=
s_e\,\mathcal{R}_e\!\left[\mathbf{I}^{\rm ref}_e\right],
\qquad
s_e \equiv 10^{\left(Z^{\rm sci}_e-Z^{\rm ref}_e\right)/2.5},
\label{eq:image_ref_align}
\end{equation}
where \(\mathcal{R}_e\) denotes the alignment of the reference image to the science frame, and \(Z^{\rm sci}_e\) and \(Z^{\rm ref}_e\) are the corresponding photometric zero points. After this step the reference image represents the static scene as it would be measured in the science-image astrometric and photometric system.

The transient itself is represented by the exposure-specific science-image \ac{PSF}. We normalize that \ac{PSF} to unit sum and shift it to the exact source position, so that the injected amplitude is controlled entirely by the source flux and the source is not artificially snapped to the nearest pixel center. The apparent AB magnitude \(m_{e,\star}\) prescribed for epoch \(e\) is converted to image-space flux through the science-image zero point:
\begin{equation}
F^{\rm inj}_e
=
10^{-0.4\left(m_{e,\star}-Z^{\rm sci}_e\right)}.
\label{eq:image_mag_to_flux}
\end{equation}
The injected science image is then
\begin{equation}
\mathbf{I}^{\rm sci}_{e,\mathrm{inj}}[u,v]
=
\mathbf{I}^{\rm sci}_e[u,v]
+
F^{\rm inj}_e\,\widehat{P}_{e,\star}[u-u_\star,v-v_\star],
\label{eq:image_injection}
\end{equation}
where \(\widehat{P}_{e,\star}\) denotes the shifted, normalized \ac{PSF}. Optional source Poisson noise and detector read noise can be added at this stage, but the dominant background, host structure, and crowding are already present in the archival science image. The reference image is not modified, because the injected transient is absent by construction from the static template.

\begin{figure*}[t]
  \centering
  \includegraphics[width=\textwidth]{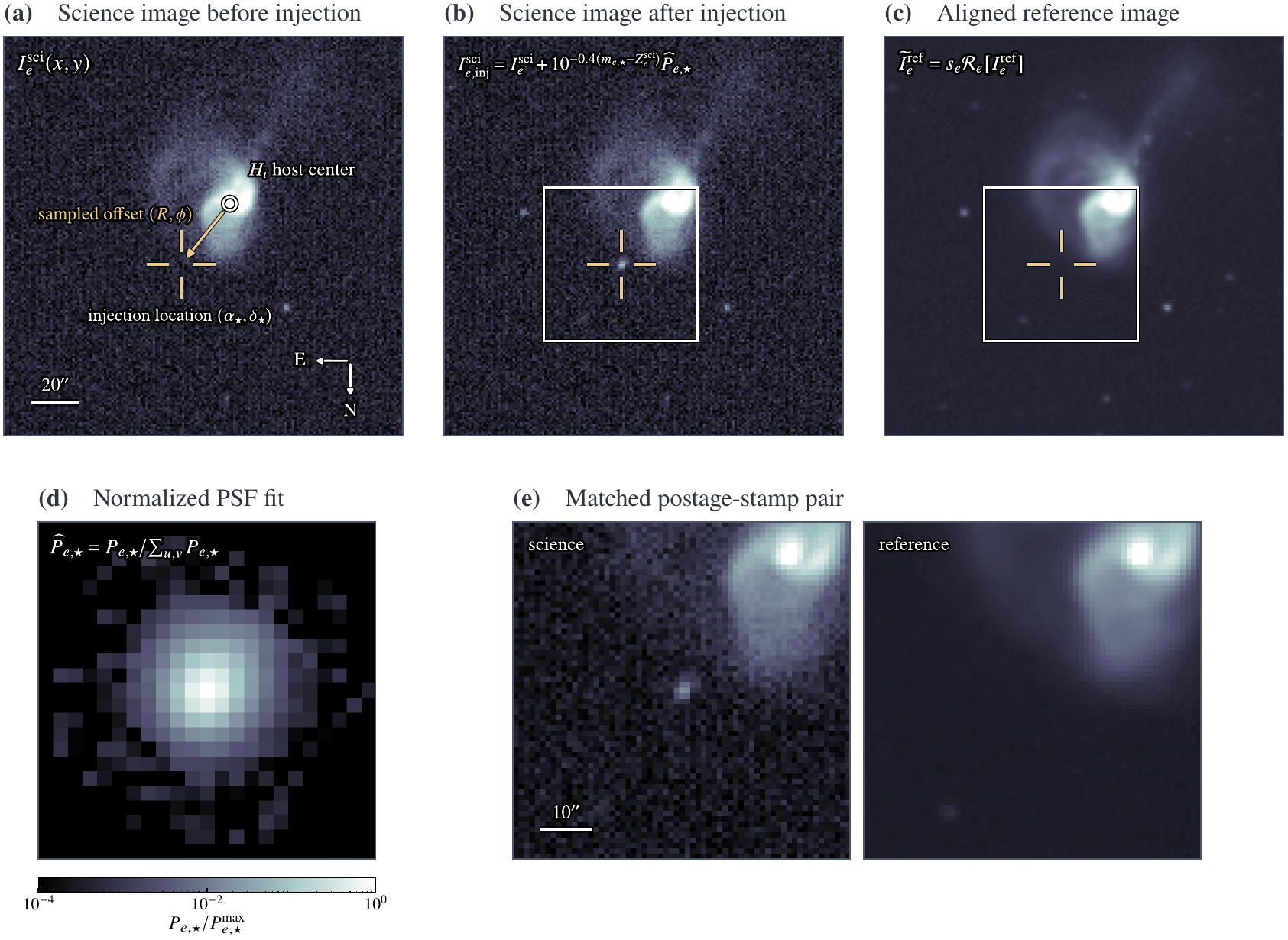}
  \caption{
Single-epoch construction of one host-associated, matched \ac{ZTF} science/reference postage-stamp pair. Panel~(a) shows the selected archival science exposure before injection. The circular marker identifies the host center, the gold vector represents the sampled projected offset $(R,\phi)$, and its endpoint defines the injection coordinates $(\alpha_\star,\delta_\star)$. The compass and 20-arcsec scale are derived from the science-image astrometric solution. Panel~(b) shows the same exposure after adding the shifted, unit-normalized exposure PSF with the flux implied by the prescribed AB magnitude,
$I_{e,\rm inj}^{\rm sci}=I_e^{\rm sci}+10^{-0.4(m_{e,\star}-Z_e^{\rm sci})}\widehat P_{e,\star}$.
Panel~(c) shows the corresponding archival reference after astrometric alignment and transfer to the science-image flux scale,
$\widetilde I_e^{\rm ref}=s_e\mathcal R_e[I_e^{\rm ref}]$,
where $s_e=10^{(Z_e^{\rm sci}-Z_e^{\rm ref})/2.5}$; the transient is added only to the science exposure. In panels~(b) and~(c), the crosshair marks the common sky coordinate and the white square marks the common $63\times63$-pixel crop. Panel~(d) displays the normalized exposure PSF used for the injection, with intensity shown relative to its maximum. Panel~(e) contains the resulting $63\times63$-pixel science and reference stamps, both centered at $(\alpha_\star,\delta_\star)$. 
}
  \label{fig:image_operator}
\end{figure*}

For downstream use, the aligned science/reference pair is cropped to a fixed square stamp:
\begin{equation}
\left(\mathbf{S}^{\rm sci}_e,\mathbf{S}^{\rm ref}_e\right)
\equiv
\left(
\mathcal{C}_L\!\left[\mathbf{I}^{\rm sci}_{e,\mathrm{inj}};\mathbf{c}_e\right],
\mathcal{C}_L\!\left[\widetilde{\mathbf{I}}^{\rm ref}_e;\mathbf{c}_e\right]
\right).
\label{eq:image_stamp_pair}
\end{equation}
In our standard products \(L=63\) pixels, and \(\mathbf{c}_e\) is usually chosen to be the injected source position so that the transient remains centered across epochs. Figure~\ref{fig:image_operator} illustrates the single-epoch construction. The archival science exposure is shown before and after injection, together with the reference image aligned to the science-image astrometric frame and photometric scale. The lower panels show the resulting \(63\times63\)-pixel science and reference stamps and the normalized science-image PSF used to place the transient. The injected source occupies a fixed position relative to the host, while the surrounding galaxy structure, neighboring sources, and background are inherited from the archival exposure.

\subsection{Host-associated kilonova postage-stamp pairs}
\label{subsec:image_host_assoc}

To specialize the generic operator above to kilonovae, we must specify how the transient is placed relative to a host galaxy. Let the candidate host set be
\begin{equation}
\mathcal{H}=\{H_i\}_{i=1}^{N_{\rm host}},
\end{equation}
with each entry \(H_i\) containing at minimum a sky position \((\alpha_i,\delta_i)\), a distance or redshift, and whatever host descriptors are available for the astrophysical population model under consideration. The general problem is to assign a prior probability that a kilonova-producing compact-binary merger occurs in host \(H_i\). In practice this is most naturally expressed through a non-negative host weight \(w_i\) that encodes the relative prior expectation that host \(H_i\) contributes to the merger population, in the same spirit that host-galaxy properties are used to interpret compact-merger environments and rates \citep{LeiblerBerger2010,SafarzadehBerger2019,Fong2022HostsI,Nugent2022HostsII}. The corresponding discrete host prior is then
\begin{equation}
p(H_i)=\frac{w_i}{\sum_{j=1}^{N_{\rm host}} w_j}.
\label{eq:host_prob_practical}
\end{equation}
The detailed construction of \(w_i\) is intentionally left open at this stage. Depending on the application, it may reflect stellar population arguments, star-formation history, morphology, environment, or any externally computed proxy for merger likelihood. If a host-weight column is supplied directly, it can be used as is; if only a filtered host list is available, the simulation reduces to sampling from that list with the desired weights. The essential point is simply that the host prior carries the astrophysical information, while the image operator remains independent of how those weights were derived.

Conditional on the selected host, the transient is assigned a projected galactocentric offset. Let \(R\) denote the projected physical offset in the host plane and let \(\phi\) denote the corresponding azimuthal angle. A convenient general placement model is
\begin{equation}
p(R,\phi\mid H_i)=p_R(R\mid H_i)\,p_\phi(\phi\mid H_i),
\label{eq:offset_factorized}
\end{equation}
with the radial term \(p_R(R\mid H_i)\) describing the projected offset distribution and the angular term \(p_\phi(\phi\mid H_i)\) describing the azimuthal placement. This is the natural form when one wants to calibrate the radial part directly against observed projected offsets, as is commonly done for short-\ac{GRB} host studies \citep{FongBerger2013,Fong2022HostsI,Berger2010NoHost}. Unless one has a physically motivated anisotropic prior tied to the host morphology, the angular distribution is taken to be isotropic,
\begin{equation}
p_\phi(\phi\mid H_i)=\frac{1}{2\pi},
\qquad
0\le \phi < 2\pi.
\label{eq:offset_isotropic_phi}
\end{equation}

Once \(R\) and \(\phi\) have been drawn, the injected sky position follows from the angular-diameter distance to the host:
\begin{equation}
\Delta\vartheta_i = \frac{R}{D_A(z_i)},
\label{eq:small_angle_offset_general}
\end{equation}
which is an excellent approximation in the small-angle regime relevant to galaxy-scale offsets. Introducing local tangent-plane coordinates \((\Delta\xi_i,\Delta\eta_i)\),
\begin{equation}
\Delta\xi_i = \Delta\vartheta_i\cos\phi,
\qquad
\Delta\eta_i = \Delta\vartheta_i\sin\phi,
\label{eq:host_offset_tangent}
\end{equation}
the corresponding celestial-coordinate increments are
\begin{equation}
\Delta \alpha_i = \frac{\Delta\xi_i}{\cos\delta_i},
\qquad
\Delta \delta_i = \Delta\eta_i,
\label{eq:host_offset_sky}
\end{equation}
so that the transient position is
\begin{equation}
\alpha_\star = \alpha_i + \Delta\alpha_i,
\qquad
\delta_\star = \delta_i + \Delta\delta_i.
\end{equation}
The factor of \(\cos\delta_i\) is the usual conversion between a tangent-plane longitudinal displacement and a change in right ascension. In the nearby limit this reduces to the familiar small-angle relation \(\Delta\vartheta \simeq R/D\).

 Once the tuple \((H_i,R,\phi)\) has been realized, the image-generation machinery requires only three ingredients: the sky position \((\alpha_\star,\delta_\star)\), the exposure metadata \((t_e,b_e,Z^{\rm sci}_e,\ldots)\), and the prescribed magnitude \(m_{e,\star}\) at that epoch. The pipeline queries the archive at the injected sky position, identifies science and reference products in the requested filters and time range, and then applies the single-epoch construction of Eqs.~\eqref{eq:image_ref_align}--\eqref{eq:image_injection}. In this sense, the kilonova specificity is entirely contained in the host prior, the offset prior, and the supplied apparent-magnitude law.

\subsection{Building a multi-band triplet sequence from a model light curve}
\label{subsec:image_sequence}

We construct a temporal image sequence by applying the single-epoch injection operator repeatedly to the same simulated event. The host, projected offset, and transient sky position are fixed throughout the sequence. At each selected visit, the injected flux is prescribed by the continuous observer-frame light curve in the corresponding filter. The magnitude-to-flux conversion uses the zero point of the archival science exposure, and the source is placed using that exposure's PSF. The resulting sequence therefore combines the temporal and color evolution of one kilonova realization with the exposure-dependent imaging conditions present in the archival data.

The underlying model magnitude determines the injected source amplitude. The detection flag and limiting magnitude remain attributes of the photometric observation model. An epoch classified as a nondetection can therefore still have an associated injected image, in which the transient contributes flux below the adopted photometric threshold. Retaining such epochs allows downstream image models to encounter the fading source and its host environment even when the photometric product contains only an upper limit.

Each selected epoch yields an aligned science/reference stamp pair, accompanied by its time, filter, and exposure metadata. The reference stamp supplies the static comparison scene, while the injected science stamp contains the evolving transient. Keeping the source position fixed across epochs preserves the association between the light curve and the local host environment; using the exposure-specific PSF and photometric calibration preserves the imaging conditions of each archival observation.

\begin{figure*}[t]
  \centering
  \includegraphics[width=\textwidth]{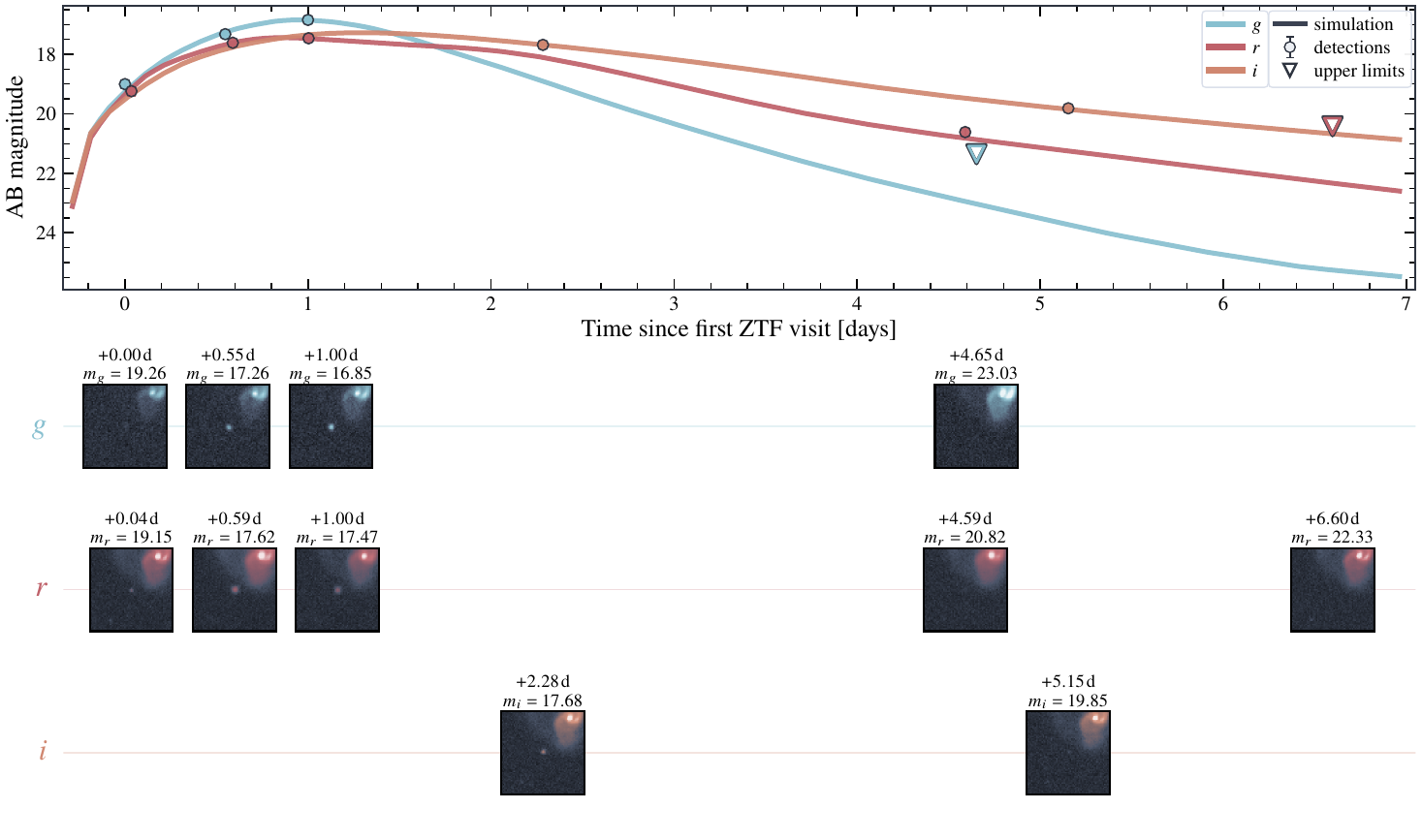}
  \caption{
    From a simulated kilonova realization to a multiband image sequence. The upper panel shows an AT2017gfo-like \ac{ZTF} realization from the LANLTP2 model family, evaluated at the adopted literature distance of NGC\,1614, \(D=64\,\mathrm{Mpc}\). Filled points with error bars denote simulated detections, open downward triangles denote simulated upper limits, and the solid curves show the underlying model light curves used to set the injected fluxes. Open squares mark the specific sampled visits used in the image construction below. The three image rows show, respectively, \(g\)-, \(r\)-, and \(i\)-band aligned reference stamps in the first column followed by injected science stamps at those selected visits, all at a fixed off-nuclear sky position relative to the host. This example illustrates how a single simulated kilonova realization is converted into a temporally ordered, filter-resolved set of realistic science/reference postage-stamp pairs while preserving the true host-galaxy morphology and sky background of the survey data.
  }
  \label{fig:gw170817_sequence}
\end{figure*}

Figure~\ref{fig:gw170817_sequence} illustrates this construction for an AT2017gfo-like LANLTP2 realization placed at an off-nuclear position in NGC\,1614, using the adopted host distance of \(64\,\mathrm{Mpc}\). The upper panel shows the underlying \(gri\) light curves together with simulated detections and upper limits. The lower rows display injected science stamps in each filter at their corresponding phases, with time measured relative to the first ZTF visit. The transient is prominent around maximum light and becomes progressively less visible as it fades, while the host remains present in every stamp.

\section{VALIDATION}

\subsection{Images}

%\subsection{Forced-photometric recovery of the injected source}

The validation measurements are performed with forced \ac{PSF} photometry at the known injected position. For an epoch \(e\), let
\begin{equation}
\mathbf{D}^{\rm inj}_e
\equiv
\mathbf{I}^{\rm sci}_{e,\mathrm{inj}}
-
\mathbf{I}^{\rm sci}_e
\label{eq:appendix_injected_only_image}
\end{equation}
denote the image containing only the injected component. By construction, Eq.~\eqref{eq:appendix_injected_only_image} removes the static host-galaxy light and sky realization already present in the archival science image, so that the recovery test isolates the photometric fidelity of the injection operator itself.

At the fixed source position \((x_\star,y_\star)\), we adopt the same sub-pixel shifted, unit-normalized \ac{PSF} model \(P_{e,\star}(u,v)\) used in the forward injection step. The forced-photometric flux estimate is then obtained as the weighted least-squares amplitude of that \ac{PSF} model on the injected-only image,
\begin{equation}
\widehat{F}_e
=
\frac{
\sum\limits_{u,v}
P_{e,\star}(u,v)\,
D^{\rm inj}_e(u,v)\,
V_e^{-1}(u,v)
}{
\sum\limits_{u,v}
P_{e,\star}^2(u,v)\,
V_e^{-1}(u,v)
},
\label{eq:appendix_forced_flux}
\end{equation}
where \(V_e(u,v)\) is the per-pixel variance model over the fitting stamp. Equation~\eqref{eq:appendix_forced_flux} is the usual maximum-likelihood estimator for the amplitude of a known \ac{PSF} template under independent Gaussian pixel noise. The corresponding formal flux variance is
\begin{equation}
\sigma^2_{\widehat{F}_e}
=
\left[
\sum\limits_{u,v}
P_{e,\star}^2(u,v)\,
V_e^{-1}(u,v)
\right]^{-1},
\label{eq:appendix_forced_flux_variance}
\end{equation}
and the associated signal-to-noise ratio is \(\mathrm{SNR}_e=\widehat{F}_e/\sigma_{\widehat{F}_e}\). In the present validation implementation we do not propagate a full empirical variance plane through the injected-only image, so the recovery step adopts a constant variance \(V_e(u,v)=\widehat{\sigma}_{e,\rm bkg}^2\), where \(\widehat{\sigma}_{e,\rm bkg}\) is estimated robustly from the local background of \(\mathbf{D}^{\rm inj}_e\) through a median-absolute-deviation scale estimator.

Whenever \(\widehat{F}_e>0\), the recovered magnitude is obtained by mapping the fitted flux back through the science-image zero point,
\begin{equation}
m_{e,\rm rec}
=
Z^{\rm sci}_e
-2.5\log_{10}\widehat{F}_e,
\label{eq:appendix_forced_mag}
\end{equation}
with the usual small-error propagation
\begin{equation}
\sigma_{m,e}
\simeq
\frac{2.5}{\ln 10}\,
\frac{\sigma_{\widehat{F}_e}}{\widehat{F}_e}.
\label{eq:appendix_forced_magerr}
\end{equation}
The validation residuals quoted below are therefore direct tests of whether the forward image injection and the backward forced-photometric recovery return consistent fluxes when evaluated on the same real survey images.

\subsubsection{Injection recovery on real science images}

Our first test checks the image-generation operator against itself. We draw host/exposure configurations from the same archive query used in production, evaluate the science-image \ac{PSF} at the injection position, and place point sources with apparent magnitudes sampled continuously over \(18 \le m_{\rm inj} \le 22\) mag. Recovery is then carried out with the forced-photometric estimator of Eqs.~\eqref{eq:appendix_forced_flux}--\eqref{eq:appendix_forced_mag} on the injected-only image of Eq.~\eqref{eq:appendix_injected_only_image}. This tests the magnitude-to-flux conversion, the sub-pixel \ac{PSF} placement, the local stamp extraction, and the behavior of the injected source on top of the real sky and host-galaxy background.

We adopt a Short-GRB-informed nearby-host selection for performing the injections. The most direct empirical guide to the environments of binary-neutron-star mergers is the observed short-\ac{GRB} host population. Short-\ac{GRB} hosts span both star-forming and quiescent galaxies \citep{Fong2013Demographics,Nugent2022HostsII}; this immediately argues against restricting kilonova image generation to a single morphological class. Their projected offsets are likewise not confined to galaxy centers. The median projected offset is \(4.5\) kpc in the sample of \citet{FongBerger2013}, while the larger compilation of \citet{Fong2022HostsI} gives median projected offsets of \(4.9\) kpc for the robust host-association sample and \(7.9\) kpc for the full sample. In both cases there is a substantial large-offset tail, including events beyond \(10\) kpc and apparently hostless systems \citep{Berger2010NoHost,FongBerger2013,Fong2022HostsI}. These measurements motivate both the host prior and the offset prior used for kilonova image generation.

For this sample, we assemble a compact proposal catalog of \(64\) nearby galaxies spanning \(31.1 \le D_L \le 85.4~\mathrm{Mpc}\). Each entry has a measured sky position and distance or redshift, together with usable archival \ac{ZTF} science and reference imaging in \(g\) and \(r\). The set includes spirals, disturbed star-forming systems, merger remnants, and quiescent early-type galaxies, thereby retaining the broad host diversity measured for short-\ac{GRB} hosts. This catalog provides an explicit finite set on which Eq.~\eqref{eq:host_prob_practical} can be evaluated. The NGC\,1614 realization shown in Figure~\ref{fig:gw170817_sequence} is one worked member of this set. Its role is to demonstrate the complete host-associated image construction; population studies can replace this proposal set and its weights without changing the image operator.

To place the transient within those hosts, we adopt an offset distribution chosen to reproduce the two most salient projected-offset diagnostics of the observed short-\ac{GRB} population: a median offset near \(5\) kpc and an appreciable large-offset tail. We represent these constraints with the two-component mixture
\begin{equation}
p_R(R)
=
\left(1-f_{\rm tail}\right)\,
\mathcal{LN}\!\left(R;\ln R_{\rm med},\sigma_{\ln R}\right)
+
f_{\rm tail}\,p_{\rm tail}(R),
\label{eq:sgrb_offset_prior}
\end{equation}
where \(\mathcal{LN}(R;\ln R_{\rm med},\sigma_{\ln R})\) is the lognormal density with median \(R_{\rm med}\), and the tail component is taken to be log-uniform,
\begin{equation}
p_{\rm tail}(R)
=
\begin{cases}
\left[R\ln\!\left(R_{\max}/R_{\min}\right)\right]^{-1}, & R_{\min}\le R\le R_{\max},\\[4pt]
0, & \text{otherwise}.
\end{cases}
\label{eq:sgrb_offset_tail}
\end{equation}
In order to match the short-\ac{GRB} offset phenomenology, we take
\begin{equation}
\begin{aligned}
R_{\rm med}&=4.2~{\rm kpc},\qquad
\sigma_{\ln R}=0.55,\qquad
f_{\rm tail}=0.20,\\
R_{\min}&=10~{\rm kpc},\qquad
R_{\max}=30~{\rm kpc},
\end{aligned}
\end{equation}
which yields an overall population median of approximately \(5.0\) kpc and \(P(R>10~{\rm kpc})\simeq 0.246\).

With this selection of hosts, the validation sample contains \(2{,}304\) injections distributed across \(64\) host fields. For each host, we use one real science exposure in each of \(g\), \(r\), and \(i\), and inject \(12\) sources spanning the adopted magnitude interval into every filter image. The resulting sample therefore contains \(768\) realizations per filter. Figure~\ref{fig:image_pipeline_validation}(c) shows the residuals in each band,
\(\Delta m \equiv m_{\rm rec}-m_{\rm inj}\),
as a function of injected magnitude. The colored curves trace the running median in each band, and the shaded regions enclose the \(16\)th--\(84\)th percentiles. Over the full sample, we find
\(\mathrm{median}\,|\Delta m|=0.023\) mag and
\(\mathrm{RMS}(\Delta m)=0.051\) mag.
For injections brighter than \(20\) mag, the recovery remains tight, with
\(\mathrm{median}\,|\Delta m|=0.015\) mag and
\(\mathrm{RMS}(\Delta m)=0.025\) mag.
The scatter grows toward the faint end, reaching
\(\mathrm{median}\,|\Delta m|=0.060\) mag for
\(21 \leq m_{\rm inj} \leq 22\) mag.
The \(g\) and \(r\) curves remain close to zero through most of the magnitude range, while the \(i\)-band envelope is visibly broader, with
\(\mathrm{median}\,|\Delta m|=0.036\) mag and
\(\mathrm{RMS}(\Delta m)=0.063\) mag.

The archival science image already contains its realized sky background, host-galaxy light, and detector noise. The injection therefore adds only the Poisson fluctuation associated with the artificial source. For a detector gain \(g_e\), expressed in electrons per data number, and a science-image zero point \(Z_e^{\rm sci}\), the expected number of injected source electrons is
\begin{equation}
N_{e,\star}
=
g_e\,10^{-0.4\left(m_{\rm inj}-Z_e^{\rm sci}\right)}.
\label{eq:injection_source_electrons}
\end{equation}
The corresponding source-counting contribution to the magnitude uncertainty scales approximately as
\(\sigma_m \simeq (2.5/\ln 10)\,N_{e,\star}^{-1/2}\).
The median zero points of the sampled configurations are \(26.14\), \(26.23\), and \(25.46\) mag in \(g\), \(r\), and \(i\), respectively. At fixed \(m_{\rm inj}\), the \(i\)-band realizations consequently contain fewer source electrons and exhibit a broader Poisson residual distribution.

% \begin{figure*}[t]
%   \centering
%   \includegraphics[width=\textwidth]{figures/fig8_image_generation_validation_continuous.pdf}
%   \caption{
%     Residuals from the continuum injection--recovery test on real \ac{ZTF} science images. The colored curves show the running median residual in \(g\), \(r\), and \(i\), and the matching shaded regions mark the \(16\)th--\(84\)th percentiles. The sample contains \(192\) injections over \(12\) science-image configurations, with \(64\) realizations per filter. For the full set we obtain \(\mathrm{median}\,|\Delta m|=0.032\) mag and \(\mathrm{RMS}(\Delta m)=0.094\) mag. The \(i\)-band sequence is broader than \(g\) and \(r\), while all three medians remain close to zero over most of the magnitude range.
%   }
%   \label{fig:image_pipeline_validation_continuum}
% \end{figure*}

\subsubsection{Replay of a real transient light curve}

We perform a second validation using the observed \ac{ZTF} light curve of ZTF18abfhaji as the empirical input sequence. Its light curve contains \(37\) usable detections across \(g\) and \(r\), covering the rise, maximum, and decline over approximately \(60\) d. We select \(10\) detections at approximately uniform temporal quantiles within each filter, giving five replay epochs in \(g\) and five in \(r\). For each selected detection we identify an archival science exposure in the same filter and require temporal agreement at the level of \(|\Delta t|<0.03\) d. The replayed source is injected at a nearby position offset by \(8^{\prime\prime}\) from the cataloged source coordinates, which prevents flux from the original transient entering the recovery measurement while retaining the exposure-specific background, zero point, and \ac{PSF}.

\begin{figure*}[t]
  \centering
  \includegraphics[width=\textwidth]{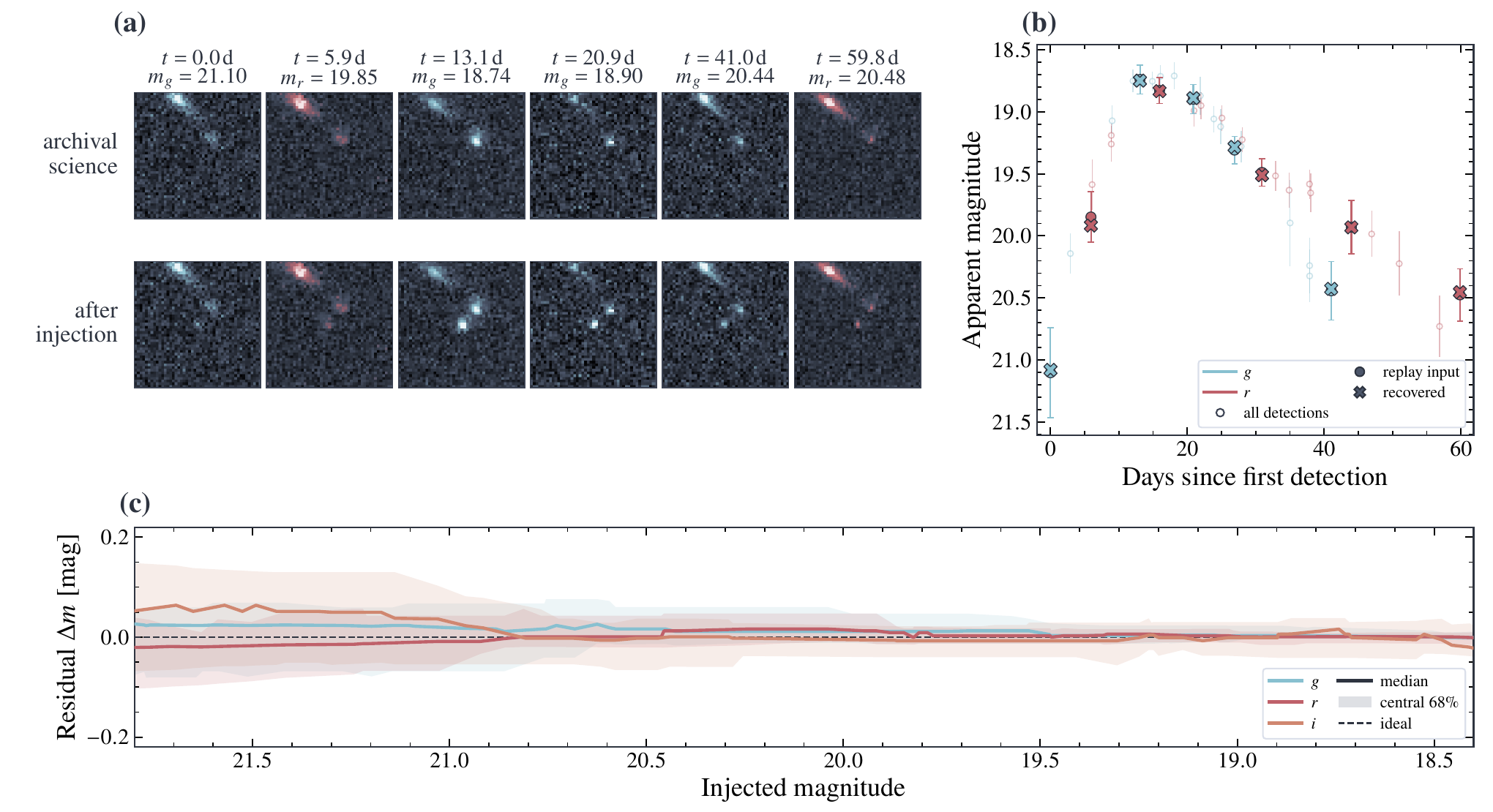}
  \caption{
    Validation of the image-injection operator on real \ac{ZTF} science exposures. Panel~(a) shows six representative epochs from the ZTF18abfhaji replay. The upper row contains the archival science stamps before injection, and the lower row contains the same scenes after adding the source. Headers give the time since the first detection and the prescribed band-specific apparent magnitude, \(m_g\) or \(m_r\). Each stamp is displayed with an inverse-hyperbolic-sine stretch referenced to the robust sky noise. The sequential maps share a neutral low-surface-brightness range and transition toward the corresponding \(g\)- or \(r\)-band color for significant positive flux. One sky estimate and one display range are applied to each before-and-after pair, preserving their direct surface-brightness comparison. Panel~(b) shows the complete observed \(g\)- and \(r\)-band light curve as open points. Filled circles identify the ten filter-balanced replay inputs, and crosses show the corresponding recovered magnitudes after injection at a position offset by \(8^{\prime\prime}\) from the cataloged transient. The replay gives \(\mathrm{median}\,|\Delta m|=0.019\) mag and \(\mathrm{RMS}(\Delta m)=0.027\) mag. Panel~(c) shows the continuum injection-recovery test for \(2{,}304\) sources distributed across \(64\) host fields. Each host contributes \(12\) injections in each of \(g\), \(r\), and \(i\), giving \(768\) realizations per filter.https://umn.zoom.us/j/8070507201 Curves give the running median of \(\Delta m=m_{\rm rec}-m_{\rm inj}\), and the shaded regions contain the central \(68\%\) of each local residual distribution. The full continuum sample has \(\mathrm{median}\,|\Delta m|=0.023\) mag and \(\mathrm{RMS}(\Delta m)=0.051\) mag. The \(i\)-band sequence is broader because its sampled zero points yield fewer source electrons at fixed apparent magnitude.
  }
  \label{fig:image_pipeline_validation}
\end{figure*}

The recovered magnitudes reproduce the replay inputs with \(\mathrm{median}\,|\Delta m| = 0.019\) mag and \(\mathrm{RMS}(\Delta m)=0.027\) mag. Figure~\ref{fig:image_pipeline_validation}(a) shows six representative archival science stamps and their post-injection realizations. Figure~\ref{fig:image_pipeline_validation}(b) places all ten replay measurements on the full observed light curve. The individual residuals lie in the interval \(-0.025 \le \Delta m \le 0.069\) mag, with no monotonic drift across the sampled phase range. This agreement verifies that exposure-specific zero points and PSFs preserve the prescribed time-dependent flux when the operator is applied across independent visits and both filters.

Taken together, these tests show few-hundredths-of-a-magnitude recovery for bright and moderate-\ac{SNR} sources, while the full \(18 \le m_{\rm inj} \le 22\) mag continuum experiment has an RMS of \(0.051\) mag. The widening residual distribution at the faint end follows the \(N_{e,\star}^{-1/2}\) dependence of source-counting noise. The pre-existing sky and host realization cancels in \(\mathbf{D}^{\rm inj}_e\), so this closure test directly validates the astrometric placement, \ac{PSF} normalization, AB magnitude conversion, and numerical stamp construction. Additional effects introduced by image subtraction, imperfect \ac{PSF} estimation, correlated backgrounds, and alert-selection thresholds remain associated with the downstream survey processing.

\subsection{Photometry}
\label{sec:photometry_validation}

In order to validate the photometric predictions from our kilonova light curve emulator, we evaluate the model on a different set of light curves presented in ~\citet{Ristic2022} that were not a part of the original training grid. Same as the original grid, these points can be found in \cite{wollaeger_2021_5745556}.
We compare the light curves predicted by the emulator to the truth light curves at the same intrinsic parameters to evaluate performance on points within the parameter space not seen during training. This section evaluates how well the model generalizes across the parameter space.

\begin{figure*}[t]
  \centering
  \includegraphics[width=0.92\textwidth]{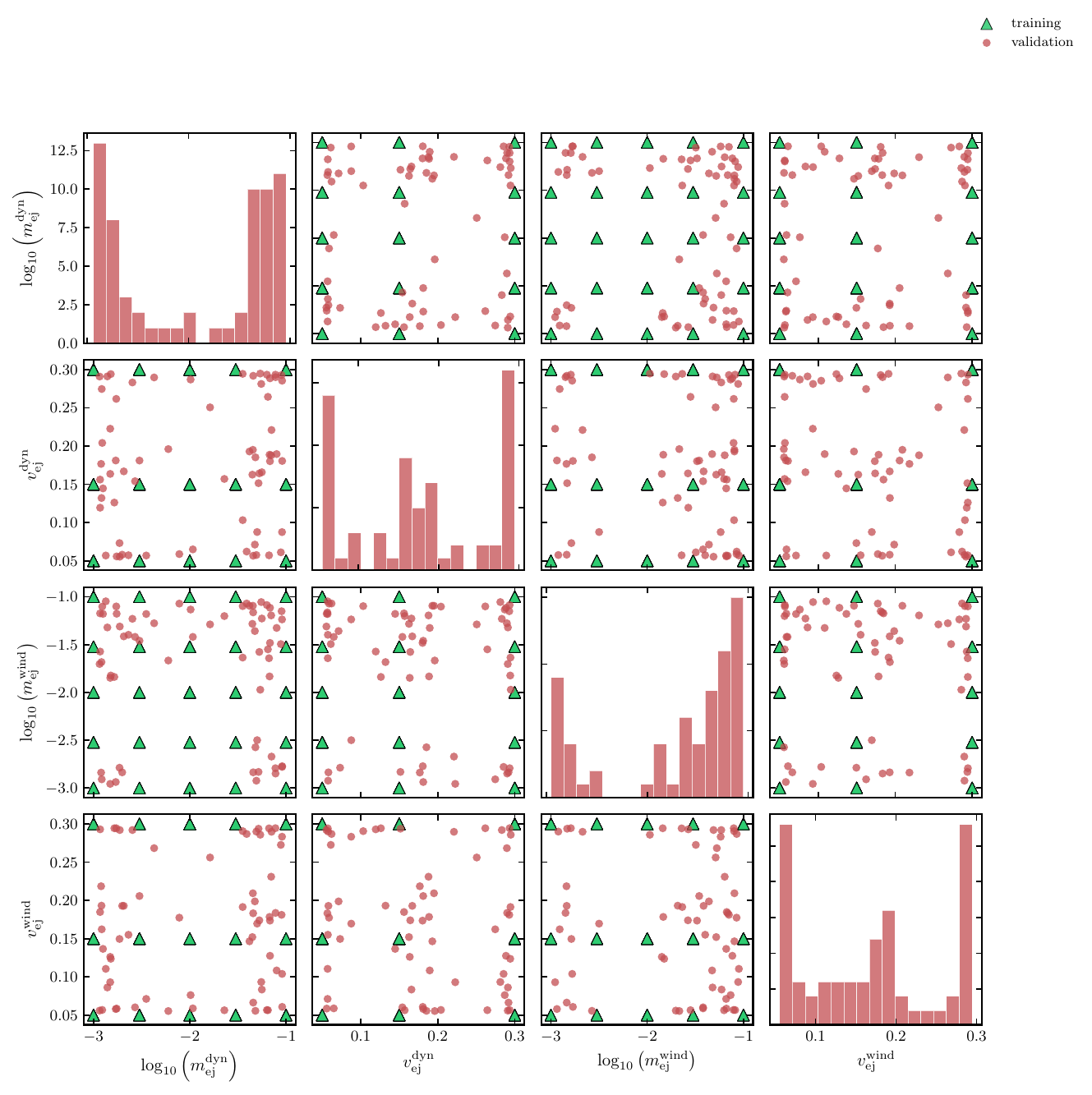}
  \caption{Coverage of the four-dimensional intrinsic-parameter space used for the photometric-surrogate evaluation. Green triangles mark the original training-grid nodes, and red circles mark the 66 validation nodes. All of the validation nodes are located to the interior of a 2\% margin on the edge of the parameter space to avoid edge cases. Off-diagonal panels show pairwise projections in \(\log_{10}M_{\rm ej}^{\rm dyn}\), \(v_{\rm ej}^{\rm dyn}\), \(\log_{10}M_{\rm ej}^{\rm wind}\), and \(v_{\rm ej}^{\rm wind}\). Diagonal panels show the marginal distributions of the selected configurations. Viewing angle is not shown here; each selected configuration is subsequently evaluated at the 54 available viewing angles.
  %All selected nodes lie strictly inside the training grid's boundary.
  }
  \label{fig:val_grid}
\end{figure*}

In Figure~\ref{fig:val_grid}, we show the original grid points and the 66 new grid points used. Each of these grid points is also evaluated on a range of $\iota$ values, specifically 54 evenly spaced bins between 0$^\circ$ and 180$^\circ$. Of the additional simulations available for validation, we select the points on the interior of the grid, as to not directly overlap with existing points or validate on the boundary of our parameter space.

%\citep{wollaeger_2021_5745556}

\begin{figure*}[t]
  \centering
  \includegraphics[width=\textwidth]{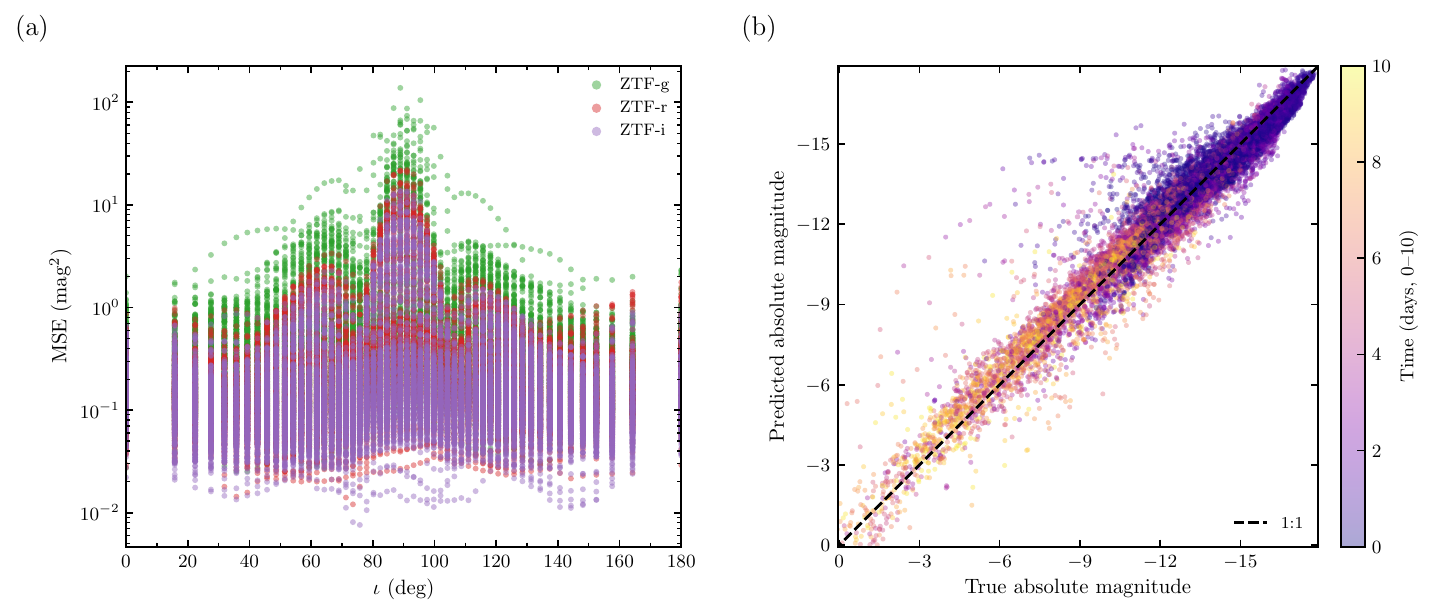}
  \caption{
  Photometric-surrogate accuracy in the three \ac{ZTF} bands ($g$, $r$, $i$), restricted to the first 10 days after explosion. \textbf{(a)} Mean squared error between predicted and true AB magnitude versus viewing angle $\iota$, evaluated over all available nodes at 54 viewing angles. Accuracy degrades sharply near edge-on viewing angles ($\iota\approx89^\circ$--$91^\circ$). \textbf{(b)} Predicted versus true absolute magnitude for the 66 selected interior configurations of Figure~\ref{fig:val_grid}, evaluated at all 54 available viewing angles, colored by time since explosion; the dashed line marks perfect agreement. Among these 66 configurations, the one with the largest mean MSE, averaged across the three \ac{ZTF} bands, reaches $53.2$~mag$^2$ at $\iota\approx88.9^\circ$. Axis limits are cropped to the $0.5$--$99.5$ percentile range of the combined predicted and true magnitude distribution to exclude a small number of numerically degenerate points at near-zero flux. Over this selected sample we find $\mathrm{median}\,|\Delta m_{\rm surr}|=0.323$~mag and $\mathrm{RMS}(\Delta m_{\rm surr})=0.763$~mag.
  }
  \label{fig:val_theta}
\end{figure*}

Figure~\ref{fig:val_theta} shows how surrogate accuracy depends on viewing angle and on the agreement between predicted and true magnitude across the validation set. Panel (a) shows the mean squared error (MSE) between predicted and true AB magnitude as a function of viewing angle $\iota$, for each of the three \ac{ZTF} bands, evaluated over all available configurations at all 54 viewing angles and restricted to the first 10 days after merger. The errors increase substantially near equatorial viewing angles, \(\iota\approx90^\circ\). Appendix~\ref{sec:appendix_high_mse} examines this near-equatorial region further, with representative light curves and ejecta parameters for the eight configurations with the highest mean MSE. The equatorial viewing angles combined with low mass ejecta values produce the faintest of light curves, and that is where our simulations break down. It is, however, unlikely a kilonova in this parameter space would be detectable. Panel (b) shows the corresponding predicted-versus-true absolute-magnitude scatter for the 66 selected validation configurations, colored by time since explosion; the dashed line marks perfect agreement. Over this selected sample we find $\mathrm{median}\,|\Delta m_{\rm surr}|=0.323$~mag and $\mathrm{RMS}(\Delta m_{\rm surr})=0.763$~mag, where $\Delta m_{\rm surr}\equiv m_{\rm pred}-m_{\rm true}$. 

\begin{figure}[t]
  %\centering
  \includegraphics[width=\columnwidth]{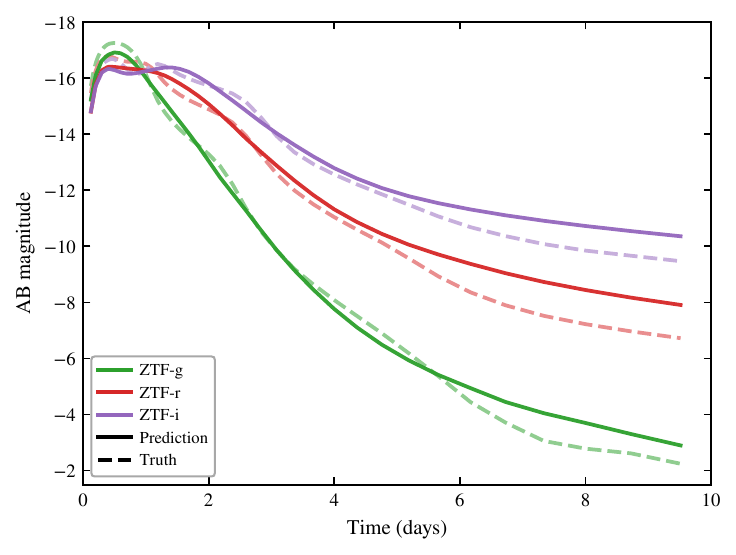}
  \caption{ Surrogate-predicted (solid) and \texttt{SuperNu} (dashed) absolute-magnitude light curves for the median MSE intrinsic-parameter--viewing-angle pair from the 66 selected configurations shown in Figure~\ref{fig:val_grid}. This example represents median performance according to the adopted curve-level MSE ranking; it is not a median location in the physical parameter space.
  }
  \label{fig:val_lightcurve_examples}
\end{figure}

Figure~\ref{fig:val_lightcurve_examples} shows a representative light curve comparison illustrating the median-level agreement reported above. To select this example, we score each intrinsic-parameter--viewing-angle pair in the selected subset by its mean squared error averaged across the three \ac{ZTF} bands over the first 10 days, sort all pairs by this score, and select the middle-ranked pair; this is distinct from the pointwise median absolute residual quoted above. Solid curves show the surrogate prediction and dashed curves show the true \texttt{SuperNu} light curve, in matching colors for each of the three \ac{ZTF} bands. 

\subsection{Spectra}

\label{sec:spectral_emulator_validation}

% \sd{do we need a subsection here?}

We validate the emulator against LANLTP2 simulations that were excluded at the level of the complete ejecta configuration.  This grouping is required because the 66 epochs and 27 viewing directions generated by one radiative-transfer simulation share the same ejecta structure and form a highly correlated sequence.  A split performed on individual spectra would place different projections of the same physical calculation in fitting and evaluation samples.  We assign the 637 available ejecta configurations to five disjoint folds in \(\boldsymbol{\vartheta}\).  Folds 2, 3, and 4 are used to fit the emulator, fold 0 is used to select the model and training checkpoint, and fold 1 is reserved for the results in this section.  The reserved sample contains 128 previously unseen ejecta configurations and 228,096 individual spectra.  Forty-five configurations lie on the Cartesian parameter grid.  The other 83 are independent LANLTP2 calculations at irregular parameter locations, selected through active learning as presented in ~\citet{Ristic2022}, which test interpolation away from the regular grid nodes.

For each simulated spectrum and emulator prediction, we measure the normalized shape error and the integrated physical flux error
\begin{equation}
E_{\rm shape}=\sum_j|\widehat{p}_j-p_j|,
\qquad
E_{\rm flux}=\sum_j\left|\frac{\widehat{L}_{\rm bol}}{L_{\rm bol}}
\widehat{p}_j-p_j\right|.
\label{eq:emulator_validation_metrics}
\end{equation}
where \(p_j\) is defined in Equation~\eqref{eq:emulator_spectral_probability}.  The first statistic lies between zero and two and measures redistribution of normalized spectral energy.  The second includes the error in bolometric luminosity and measures the wavelength-integrated absolute flux discrepancy.  We also measure the location of the broad maximum in \(\lambda f_\lambda\), the optical to near-infrared color, and the relative luminosity error.  Logarithmic flux residuals are evaluated in bins satisfying \(f_{\lambda,j}^{\rm LANLTP2}>10^{-3}\max_k f_{\lambda,k}^{\rm LANLTP2}\) for each spectrum.  This threshold retains 67.5\% of all wavelength bins and removes Monte Carlo zeros and negligible-flux regions from a statistic that is undefined at zero flux.  The integrated errors in Equation~\eqref{eq:emulator_validation_metrics} use every wavelength bin.

Across the reserved sample, the mean \(E_{\rm shape}\) is 0.0465.  A bootstrap that resamples complete ejecta configurations gives a 95\% confidence interval from 0.0437 to 0.0494, and the 95th percentile over individual spectra is 0.1096.  The mean \(E_{\rm flux}\) is 0.0555.  The median relative luminosity error is 0.0203, and its 95th percentile is 0.0899.  Within the significant-flux mask, the median spectrum has a median absolute log-flux error of 0.0215 dex, corresponding to a flux ratio of approximately 1.05.  The median optical to near-infrared color error is 0.0123 dex, or 0.031 mag.  The median displacement of the broad \(\lambda f_\lambda\) maximum is 0.0021 dex in wavelength.  Its 95th percentile reaches 0.0741 dex because two broad maxima with similar heights can exchange order under a comparatively small flux perturbation.

\begin{figure*}[t]
  \centering
  \includegraphics[width=0.82\textwidth]{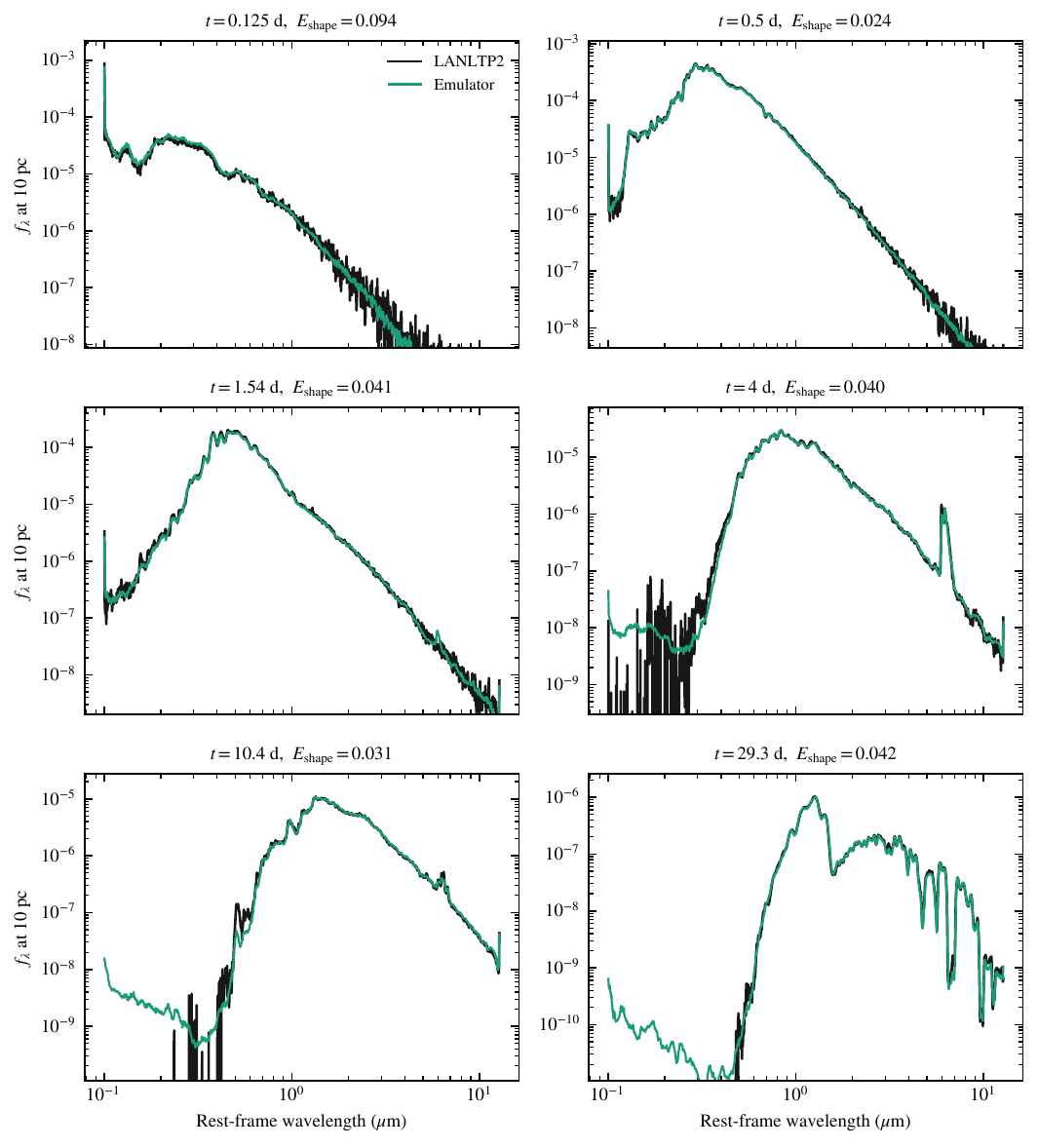}
  \caption{
     Simulated LANLTP2 spectra and emulator reconstructions for an ejecta configuration excluded from fitting and model selection.  The physical parameters are \(M_{\rm d}=0.0577\,M_\odot\), \(v_{\rm d}=0.051c\), \(M_{\rm w}=0.0218\,M_\odot\), and \(v_{\rm w}=0.116c\).  Six epochs are shown at \(\mu=0.50\).  The configuration was selected as the reserved simulation whose mean shape error lies closest to the median among all test simulations.  The simulated spectra are plotted without smoothing or renormalization.  The labels above the panels report the integrated normalized shape error from Equation~\eqref{eq:emulator_validation_metrics}.
  }
  \label{fig:emulator_spectral_evolution}
\end{figure*}

Figure~\ref{fig:emulator_spectral_evolution} follows one ejecta configuration through six epochs at \(\mu=0.50\).  We selected the configuration before plotting as the reserved simulation whose mean \(E_{\rm shape}\), averaged over all epochs and viewing directions, lies closest to the median among the 128 test simulations.  The comparison therefore represents typical model-level performance.  The emulator follows the initial blue suppression, the movement of the spectral energy distribution toward longer wavelengths, and the broad optical and near-infrared structures.  The largest difference in this sequence occurs at 0.125~d, when the spectral morphology evolves most rapidly.  At later epochs, visible discrepancies are concentrated in low-flux regions and in the detailed shapes of broad blended features.

\iffalse
\begin{figure*}[t]
  \centering
  \includegraphics[width=0.82\textwidth]{figures/wavelength_time_residuals-1.pdf}
  \caption{Population distribution of \(\log_{10}(f_\lambda^{\rm emu}/f_\lambda^{\rm LANLTP2})\) over the 228,096 spectra in the reserved test sample. Panel (a) shows the median signed residual, panel (b) shows the median absolute residual, and panel (c) shows the 90th percentile of the absolute residual as functions of rest-frame wavelength and time after merger. Bins below \(10^{-3}\) of the maximum simulated \(f_\lambda\) in each spectrum are excluded before computing the population quantiles.
  }
  \label{fig:emulator_residual_map}
\end{figure*}
\fi 

\begin{figure}[t]
  %\centering
  %\includegraphics[width=0.82\textwidth]{figures/wavelength_time_residuals-1.pdf}
  \includegraphics[width=\columnwidth]{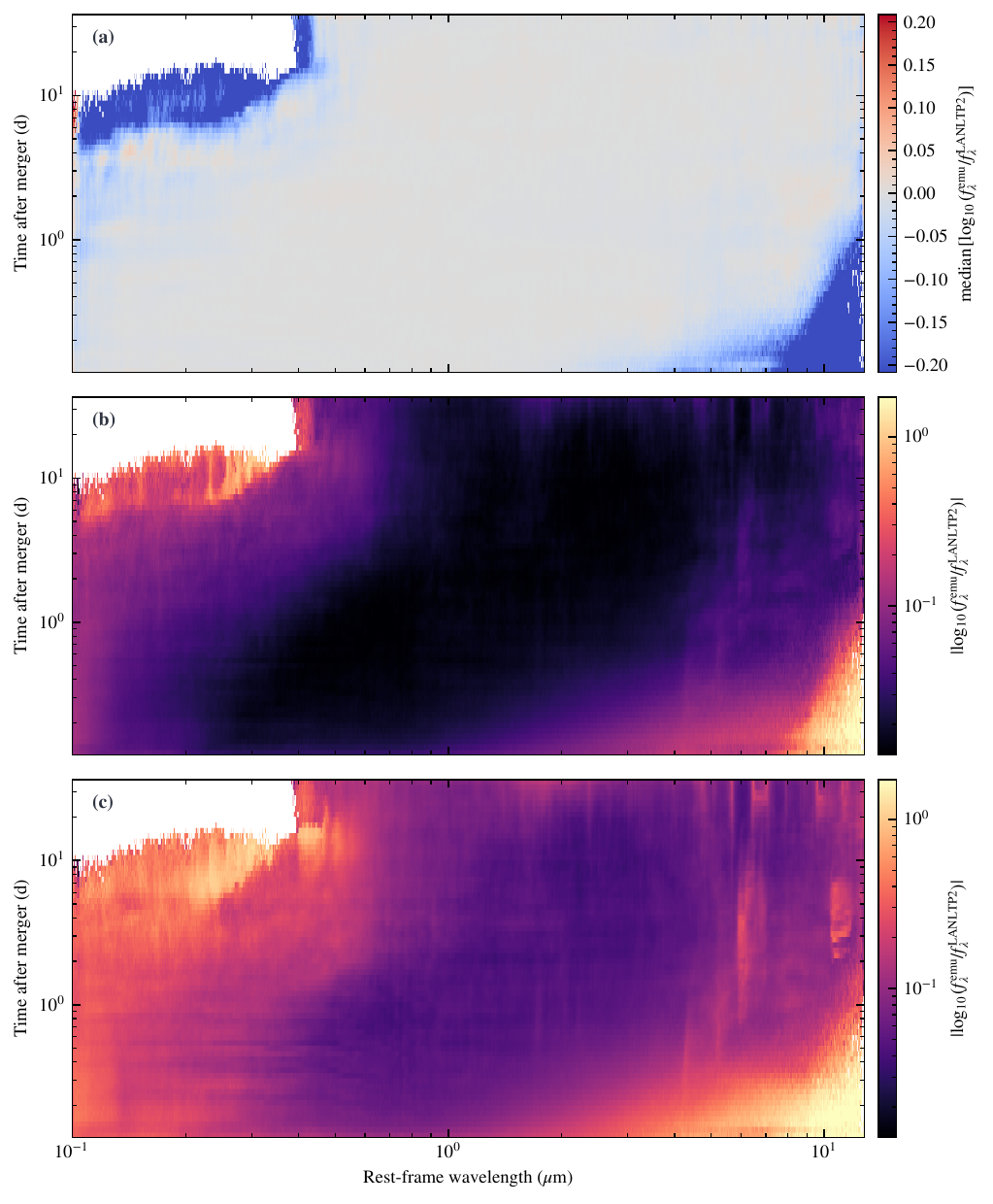}
  \caption{Population distribution of \(\log_{10}(f_\lambda^{\rm emu}/f_\lambda^{\rm LANLTP2})\) over the 228,096 spectra in the reserved test sample. Panel (a) shows the median signed residual, panel (b) shows the median absolute residual, and panel (c) shows the 90th percentile of the absolute residual as functions of rest-frame wavelength and time after merger. Bins below \(10^{-3}\) of the maximum simulated \(f_\lambda\) in each spectrum are excluded before computing the population quantiles.
  }
  \label{fig:emulator_residual_map}
\end{figure}

Figure~\ref{fig:emulator_residual_map} locates the population residuals in wavelength and time.  The median signed residual stays close to zero across most wavelength and time combinations that carry appreciable flux.  The median absolute residual is smallest through much of the optical and near-infrared evolution after the first few tenths of a day.  The early spectra and wavelength regions containing rapidly changing broad structures produce the most prominent residuals.  At 0.125~d, the median \(E_{\rm shape}\) is 0.0827 and the 90th percentile is 0.2198.  The time-dependent median reaches 0.0329 at 0.771~d and remains near that level through much of the subsequent evolution. Panel (a) identifies coherent color bias, while panels (b) and (c) separate the typical residual from the difficult upper tail.

\begin{figure*}[t]
  \centering
  \includegraphics[width=0.82\textwidth]{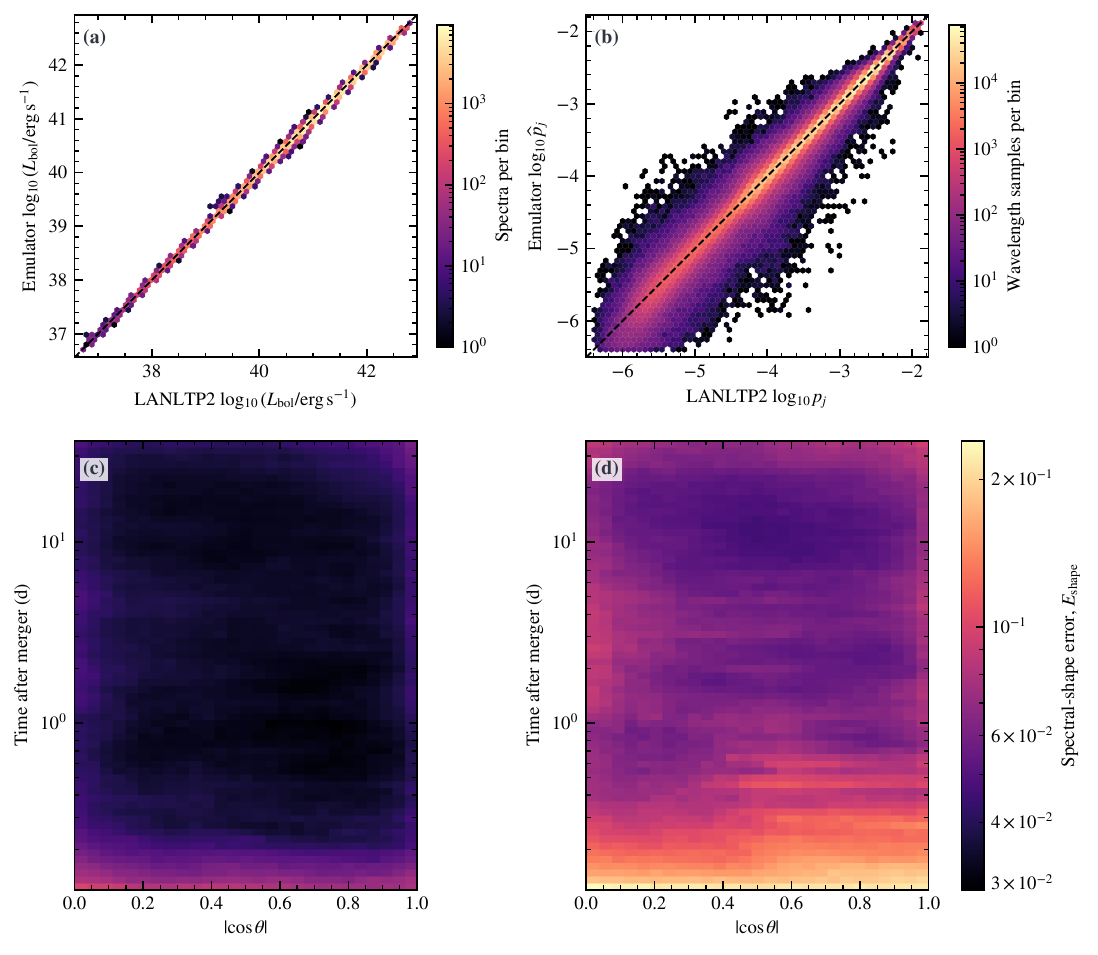}
  \caption{
    Population diagnostics for the held-out LANLTP2 simulations. Panel (a) compares emulated and simulated \(\log_{10}L_{\rm bol}\) for all 228,096 test spectra. Panel (b) compares \(\log_{10}\widehat p_j\) and \(\log_{10}p_j\) for 2,048,000 wavelength samples drawn equally from the 128 reserved ejecta configurations and restricted to bins with native \(f_\lambda\) above \(10^{-3}\) of the maximum for that spectrum. In both panels, color gives the number of samples in each hexagonal bin and the dashed line marks exact reconstruction. The displayed shape range contains the central 99.9\% of the combined native and emulated sample. Panels (c) and (d) give the median and 90th percentile of \(E_{\rm shape}\), respectively, across the 128 complete held-out ejecta configurations at each fixed combination of time and viewing angle. The largest errors occur at the earliest epochs, while the two-dimensional representation resolves the weaker angular structure and its evolution with time.
  }
  \label{fig:emulator_population_diagnostics}
\end{figure*}

The diagnostics in Figure~\ref{fig:emulator_population_diagnostics} examine the two quantities that determine the reconstructed physical flux. Panel (a) compares the emulated and native bolometric luminosities for every spectrum in the reserved sample. This quantity supplies the amplitude that converts \(\widehat p_j\) into physical flux through Equation~\eqref{eq:emulator_flux_reconstruction}. The density remains concentrated around the one-to-one relation over the full luminosity range sampled by LANLTP2.

Panel (b) directly examines the normalized spectral shape.  Each point contributing to this density is a simulated and emulated pair \((p_j,\widehat p_j)\) for the same wavelength bin of the same spectrum.  We restrict the comparison to bins satisfying \(f_{\lambda,j}^{\rm LANLTP2}>10^{-3}\max_k f_{\lambda,k}^{\rm LANLTP2}\), consistent with the significant-flux mask used for the logarithmic residuals.  We then draw 16,000 wavelength samples from each reserved ejecta configuration, with all epochs and viewing directions eligible, for a total of 2,048,000 pairs.  Equal sampling by ejecta configuration prevents a subset of physical models from dominating the density.  The logarithmic axes resolve both the high-probability spectral peaks and lower-probability structure over several orders of magnitude.  The median absolute value of \(\log_{10}(\widehat p_j/p_j)\) is 0.0202~dex, and its 90th percentile is 0.0935~dex.  In bins with \(p_j\geq10^{-3}\), which individually carry the largest fractions of spectral energy, these values decrease to 0.0126 and 0.0440~dex.  The increasing dispersion toward smaller \(p_j\) is visible in the density.  The display limits contain the central 99.9\% of the combined native and emulated sample.  

Panels (c) and (d) examine the joint dependence of \(E_{\rm shape}\) on time and viewing angle.  Each cell gives a quantile across the 128 reserved ejecta configurations at one fixed pair \((t,\mu)\).  The construction therefore preserves correlations between epoch and line of sight without treating the 66 epochs or 27 angular projections from one radiative-transfer calculation as independent members of the test population. At 0.125~d, the median varies from 0.0719 to 0.1005 across viewing angle, while the 90th percentile varies from 0.2084 to 0.2396.  Both statistics decline rapidly over the next several epochs.  Through much of the subsequent evolution, the median lies between approximately 0.03 and 0.045, with angular structure that remains smaller than the early-time degradation.  The mean model-averaged \(E_{\rm shape}\) is 0.0467 for Cartesian-grid calculations and 0.0464 for the irregularly placed simulations, which supports interpolation away from the regular grid nodes.

Taken together, the four panels show that the population mean is not produced by compensating errors between luminosity and spectral shape.  The luminosity reconstruction remains close to the identity relation, the significant-flux wavelength bins follow the corresponding identity relation in \(p_j\), and the time and viewing-angle maps locate the residual upper tail at the earliest epochs.  The agreement between the mean errors for Cartesian-grid and irregularly placed simulations also indicates that the test result is not confined to interpolation between adjacent nodes of the original regular grid.  These diagnostics assess numerical fidelity to LANLTP2.  They do not incorporate uncertainty in the radiative-transfer physics, atomic data, ejecta geometry, or composition.

\section{Comparison to GW170817}
\subsection{Light curves}
\label{subsec:gw170817_lightcurves}
\begin{figure*}[t]
  \centering
  \includegraphics[width=\textwidth]{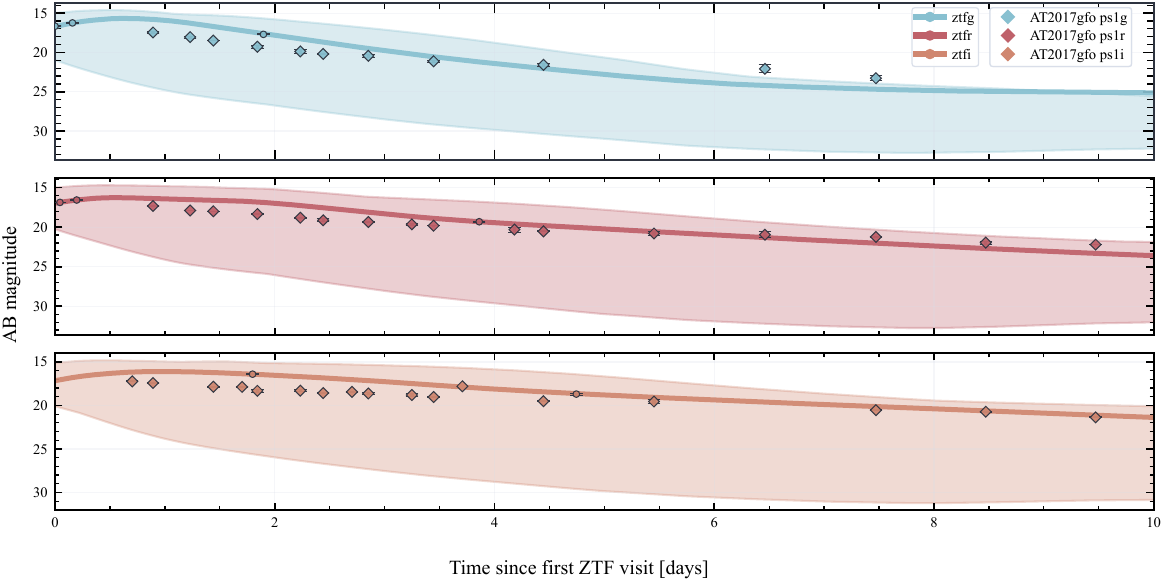}
  \caption{A comparison of AT2017gfo to simulated ZTF observations. The circle markers show the simulated ZTF observations and the solid lines are the accompanying light curves the individual observations are drawn from. The light curves are simulated using the maximum likelihood AT2017gfo parameters found in \cite{King_2025}. The shaded region corresponds to the 90\% credible interval for a set of simulated kilonova light curves as seen by ZTF within the measured distance uncertainty of GW170817. The diamond markers show the AT2017gfo observations in ps1 bands \citep{CoFo2017,Smartt:2017fuw,AbEA2017f}.}
  \label{fig:GW170817}
\end{figure*}

We now compare our simulations to observations of AT2017gfo \citep{CoFo2017,Smartt:2017fuw,AbEA2017f}, the kilonova associated with the binary neutron star merger GW170817 \citep{AbEA2017b}.
In order to do so, we start with the maximum likelihood parameters of AT2017gfo from \citet{King_2025}, and generate light curves using the above workflow, sampling realistic ZTF observations.
In addition, we simulate a varied set of kilonovae light curves within the distance uncertainty of GW170817, specifically between 26 and 48 Mpc as originally measured in \cite{AbEA2017b}.
This set is meant to show a realistic credible interval for GW170817-like neutron star mergers. Figure~\ref{fig:GW170817} shows this comparison, with circle markers and the solid lines denoting the simulated maximum likelihood light curves, the diamond markers showing AT2017gfo, and the shaded region showing the 90\% credible interval for simulated kilonovae within the measured distance uncertainty. We find the AT2017gfo observations are consistent with both the maximum likelihood and credible interval simulations, providing a sanity check of our methods.

\subsection{Spectrum}
\label{subsec:gw170817_spectra}

The time-resolved spectrum of AT2017gfo provides an external test of the spectral data-generation model. We use the public ENGRAVE reduction of the VLT/X-shooter sequence\footnote{\url{https://www.engrave-eso.org/AT2017gfo-Data-Release/}} originally presented by \citet{Pian:2017gtc} and \citet{Smartt:2017fuw}. The sequence spans 1.43 to 10.40 days after merger and covers the optical and near-infrared wavelength range. The released products have been flux calibrated, adjusted to contemporaneous photometry, corrected for Milky Way extinction, and transformed to a rest-frame wavelength coordinate. We retain their released flux-density normalization. The observed phases are converted to the source frame using \(t_{\rm rf}=t_{\rm obs}/(1+z)\), with \(z=0.009783\).

We evaluate the emulator at the maximum-likelihood LANLTP2 parameters used for the photometric comparison from \citet{King_2025}: \(M_{\rm d}=0.040\,M_\odot\), \(v_{\rm d}=0.27c\), \(M_{\rm w}=0.028\,M_\odot\), \(v_{\rm w}=0.07c\), and \(\iota=0.75\) rad. These values define the central prediction at every epoch, with \(D_L=44\) Mpc. The shaded region propagates the measured 26 to 48 Mpc distance range together with the 90\% inclination interval \(0.42\leq\iota\leq0.95\) rad reported by \citet{King_2025}. The envelope includes the changes in luminosity, color, and broad spectral morphology produced by these distance and viewing-angle ranges. No parameter is optimized against the spectra. This prescription makes the spectral sequence a predictive check of the physical realization already selected from the photometry.

\begin{figure*}[t]
  \centering
  \includegraphics[width=\textwidth,height=0.76\textheight,keepaspectratio]{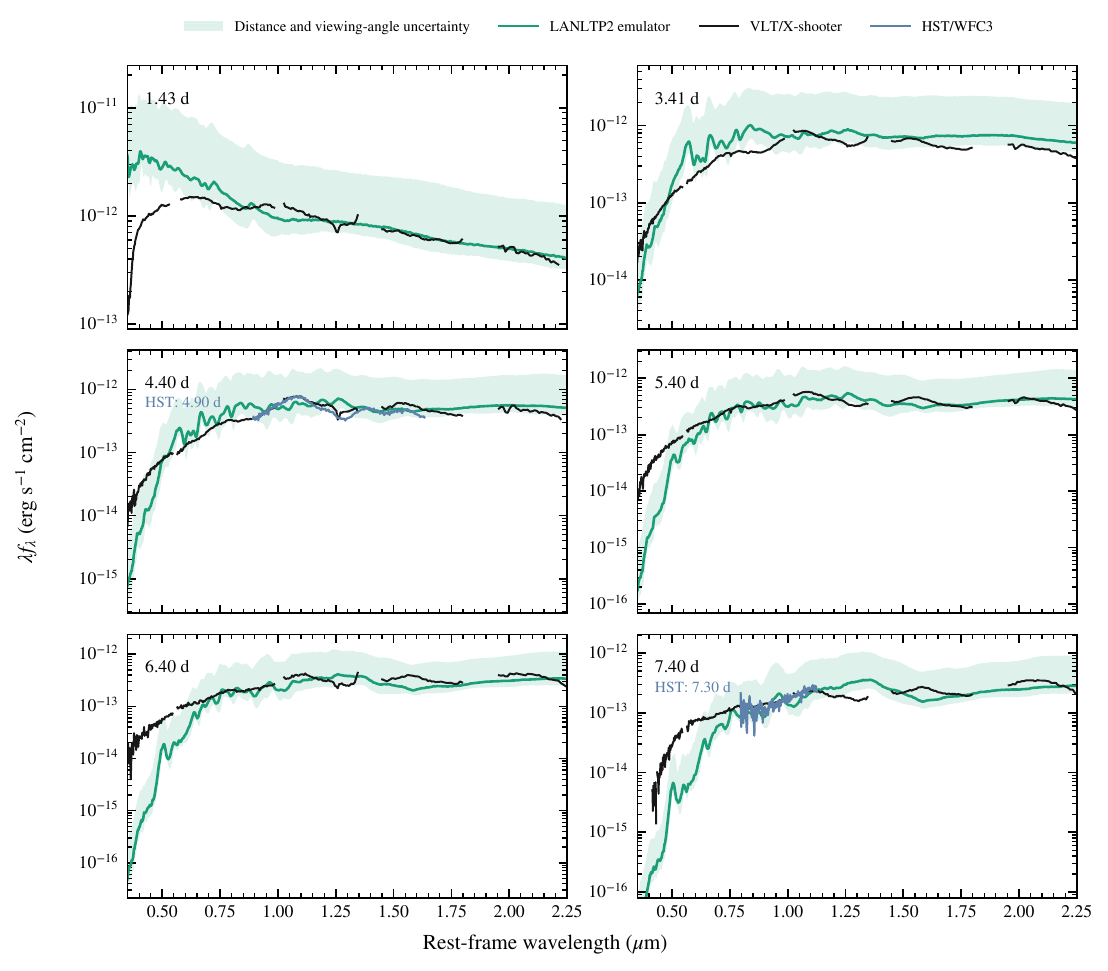}
  \caption{
    Multiepoch comparison of the LANLTP2 spectral emulator with the public spectra of AT2017gfo from 1.43 to 7.40 days after merger. Black curves show VLT/X-shooter observations at the labeled phases. Blue curves show the HST/WFC3 grism spectra at 4.9 and 7.3 days, plotted in the panels containing the nearest X-shooter epochs. Green curves show the emulator evaluated at the maximum-likelihood parameters from \citet{King_2025}, with \(\iota=0.75\) rad and \(D_L=44\) Mpc. The shaded region shows the range of simulated spectra obtained by varying the luminosity distance over the measured 26 to 48 Mpc interval and the inclination over its 90\% credible interval, \(0.42\leq\iota\leq0.95\) rad. No spectral parameters or wavelength-dependent corrections are fitted to the observations. Gaps in the X-shooter curves mark arm joins and wavelength intervals affected by strong telluric absorption.
  }
  \label{fig:at2017gfo_spectral_sequence}
\end{figure*}

Figure~\ref{fig:at2017gfo_spectral_sequence} also includes the HST G102 and G141 grism spectra at 4.9 and 7.3 days reported by \citet{Tanvir:2017pws} and distributed with the ENGRAVE release. Their integrated-flux ratios relative to the neighboring X-shooter spectra are 1.02 and 0.90 over the common wavelength intervals, with normalized shape errors of 0.092 and 0.139. The agreement supports the broad near-infrared morphology measured from the ground-based sequence. The released HST products were aligned to the contemporaneous photometry or X-shooter sequence, so they provide a cross-instrument check of spectral shape rather than an independent absolute calibration.

For a common numerical comparison, each X-shooter spectrum is rebinned onto the emulator wavelength grid by integrating the densely sampled observed flux density within each wavelength bin and dividing by the covered bin width. This flux-conserving average treats the emulator output as a bin-averaged flux density. We retain rest-frame wavelengths from 0.35 to \(2.25\,\mu\mathrm{m}\), and omit the joins between the three X-shooter arms together with the principal regions of strong telluric absorption. Let \(\mathcal{B}_t\) denote the set of retained wavelength bins at epoch \(t\). The integrated flux within this common band and its normalized spectral distribution are
\begin{equation}
F_t^X=\sum_{j\in\mathcal{B}_t}f_{\lambda,t,j}^X\Delta\lambda_j,
\qquad
p_{t,j}^X=\frac{f_{\lambda,t,j}^X\Delta\lambda_j}{F_t^X},
\label{eq:at2017gfo_band_probability}
\end{equation}
where \(X\in\{\mathrm{obs},\mathrm{emu}\}\). We summarize the absolute normalization with \(R_F(t)=F_t^{\rm emu}/F_t^{\rm obs}\), and the wavelength redistribution with
\begin{equation}
E_{\rm shape}^{\rm obs}(t)
=\sum_{j\in\mathcal{B}_t}
\left|p_{t,j}^{\rm emu}-p_{t,j}^{\rm obs}\right|.
\label{eq:at2017gfo_shape_error}
\end{equation}
This definition matches the normalized shape statistic used for held-out LANLTP2 calculations in Equation~\eqref{eq:emulator_validation_metrics}, with the sum restricted to the wavelength coverage supported by each observation.

The absolute flux of this fixed LANLTP2 realization lies close to the observed scale. At 44 Mpc, the median value of \(R_F\) is 1.064, with individual values from 0.830 to 1.931. The distance that would match the band-integrated flux at each epoch ranges from 40.1 to 61.1 Mpc. Between 5.40 and 9.40 days, the central prediction lies within 18\% of the observed integrated flux. The earliest spectra show the largest excess in total flux. The median absolute physical log-flux residual in the retained wavelength bins is 0.161 dex. The envelope in Figure~\ref{fig:at2017gfo_spectral_sequence} combines the distance dependence with the additional variation in normalization and spectral shape induced by the allowed viewing angles.

The normalized spectra agree most closely during the middle of the sequence. The median \(E_{\rm shape}^{\rm obs}\) is 0.340, and the range is 0.203 to 0.596. The median absolute logarithmic residual after normalization is 0.153 dex. From 3.41 to 6.40 days, \(E_{\rm shape}^{\rm obs}\) remains between 0.203 and 0.249. During these epochs the model follows the progressive suppression of optical emission and the movement of the broad maximum toward the near-infrared, although the location and contrast of individual structures differ. The median absolute residual in the optical to near-infrared flux ratio is 0.286 dex, corresponding to 0.72 mag. Its sign changes with time: the emulated spectrum is relatively bluer through 4.40 days and becomes increasingly red relative to the observations at later epochs.

The earliest and latest spectra expose more specific limitations. At 1.43 days, the observed broad maximum lies near \(0.62\,\mu\mathrm{m}\), while the model maximum occurs near \(0.41\,\mu\mathrm{m}\), and \(E_{\rm shape}^{\rm obs}=0.513\). After approximately six days, the observed sequence develops a broad maximum near 2.03 to \(2.10\,\mu\mathrm{m}\). Analyses of the complete AT2017gfo sequence identify strong emission-like structure near 1.58 and \(2.07\,\mu\mathrm{m}\) during these phases \citep{Gillanders:2023jpd}. The LANLTP2 prediction places its broad maximum between 1.26 and \(1.48\,\mu\mathrm{m}\) and does not reproduce the observed late-time 2.1~\(\mu\mathrm{m}\) structure. The shape error consequently rises to 0.596 at 9.40 days.

Those tensions are consistent with the discrepancies reported when the underlying radiative-transfer simulated spectra were compared directly with AT2017gfo \citep{Wollaeger:2021qgf}. After first restricting the grid using broadband information, \citet{Wollaeger:2021qgf} ranked candidate calculations through an error in bolometric luminosity. Their lowest-error example produced a broad absorption structure between 1 and \(2\,\mu\mathrm{m}\) at day 5 that was absent from the X-shooter spectrum. Models with high-\(Y_e\) ejecta velocities above \(0.05c\) were also systematically faint in the optical by that epoch, and the late-time calculations developed a spectral cliff that was not present in AT2017gfo. These examples suggest that models selected through broadband criteria can retain large, coherent wavelength-dependent errors. Consistency in bolometric luminosity does not by itself establish consistency in the spectrum shape. 

Figure~\ref{fig:at2017gfo_spectral_sequence} complements the reserved-simulation validation in Section~\ref{sec:spectral_emulator_validation}. The reserved calculations measure how faithfully the numerical emulator reconstructs the LANLTP2 radiative-transfer library. The AT2017gfo comparison measures how closely one member of that library, selected by the photometric maximum-likelihood solution, describes a real kilonova. The corrected absolute normalization is broadly consistent with the data, particularly from approximately four to nine days. The wavelength-dependent residuals remain much larger than the emulator errors measured against held-out LANLTP2 calculations, and they have a coherent temporal evolution. Their character agrees with the limitations documented for the original grid. A joint spectroscopic fit could move to a different location in the ejecta parameter space. The restricted physical assumptions of the library would remain. These sources of physical model uncertainty should be represented when synthetic spectra are used to train parameter-estimation models for observed kilonovae.

\section{Discussion}
In this work we present a pipeline for the simulation of multi-modal observations of kilonovae with ZTF. Realistic photometry, images, and spectra are generated using kilonova models and observed survey distributions. Together, these three data types are representative of the electromagnetic observations that may be available for candidate kilonovae, and constitute a dataset capable of training AI models intended for kilonova classification, identification, and parameter estimation. We validate our simulation methods and find they accurately produce realistic survey detections consisting of images, spectra, and photometry. 

We intend for our methods and simulated data to be used to predict what future kilonova observations may look like, as well as for training models, both of which may improve our chances of making the next kilonova detection. Our methods freely generalize to other observatories and surveys, especially the LSST all-sky survey conducted by the Vera C. Rubin Observatory. In the future, we will improve our methodology to produce joint observations between ZTF and Vera C. Rubin.

\begin{acknowledgments}
We thank Ingo Tews, Niharika Sravan, and Natalya Pletskova for helpful discussions. This work was supported by the U.S. Department of Energy through Los Alamos National Laboratory (LANL). 
LANL is operated by Triad National Security, LLC, for the National Nuclear Security Administration of U.S. Department of Energy (Contract No.~89233218CNA000001).
F.F.N., F.T., L.L.J. S.C., and S.D. were supported by the Laboratory Directed Research and Development program of Los Alamos National Laboratory project 20250750ECR.
A.T. acknowledges support from the Director's Postdoc Fellowship at Los Alamos National Laboratory funded by the Laboratory Directed Research and Development Program project number 20251157PRD2. 
F.F.N. and M.W.C. acknowledge support from the National Science Foundation with grant numbers PHY-2117997, PHY-2308862 and PHY-2409481.
This research used resources of the National Energy Research
Scientific Computing Center, a DOE Office of Science User Facility
supported by the Office of Science of the U.S. Department of Energy
under Contract No. DE-AC02-05CH11231 using NERSC award
NP-ERCAP0037764. 
This research also used resources provided by the Los Alamos National Laboratory Institutional Computing Program, which is supported by the U.S. Department of Energy National Nuclear Security Administration under Contract No. 89233218CNA000001.
SED Machine is based upon work supported by the
National Science Foundation under Grant No. 1106171.
The Gordon and Betty Moore Foundation, through both
the Data-Driven Investigator Program and a dedicated
grant, provided critical funding for SkyPortal. The authors used AI-assisted tools (ChatGPT 6.1 Sol and Claude Sonnet 5.5) for translation, grammar correction, and proofreading during manuscript preparation.
% LA-UR-26-24831
\end{acknowledgments}

\software{Astropy \citep{astropy:2013,astropy:2018,astropy:2022},
          CIGALE \citep{Boquien2019},
          HyperGal \citep{Lezmy2022},
          Matplotlib \citep{Hunter:2007},
          NumPy \citep{harris2020array},
          pandas \citep{mckinney-proc-scipy-2010},
          PyTorch \citep{NEURIPS2019_bdbca288},
          reproject \citep{Robitaille2018reproject},
          scikit-learn \citep{JMLR:v12:pedregosa11a},
          SciPy \citep{2020SciPy-NMeth},
          SEP \citep{Barbary2016SEP,BertinArnouts1996},
          SuperNu \citep{Wollaeger2013SuperNu,Wollaeger2014SuperNu},
          TensorFlow \citep{tensorflow2015-whitepaper}}


\appendix

%\section{Appendix information}
%Appendix 
%\rednote{mention high MSE lcs?}

\section{High-MSE light curves near equatorial viewing angles}
\label{sec:appendix_high_mse}

\begin{figure}[t]
  \centering
  \includegraphics[width=\columnwidth]{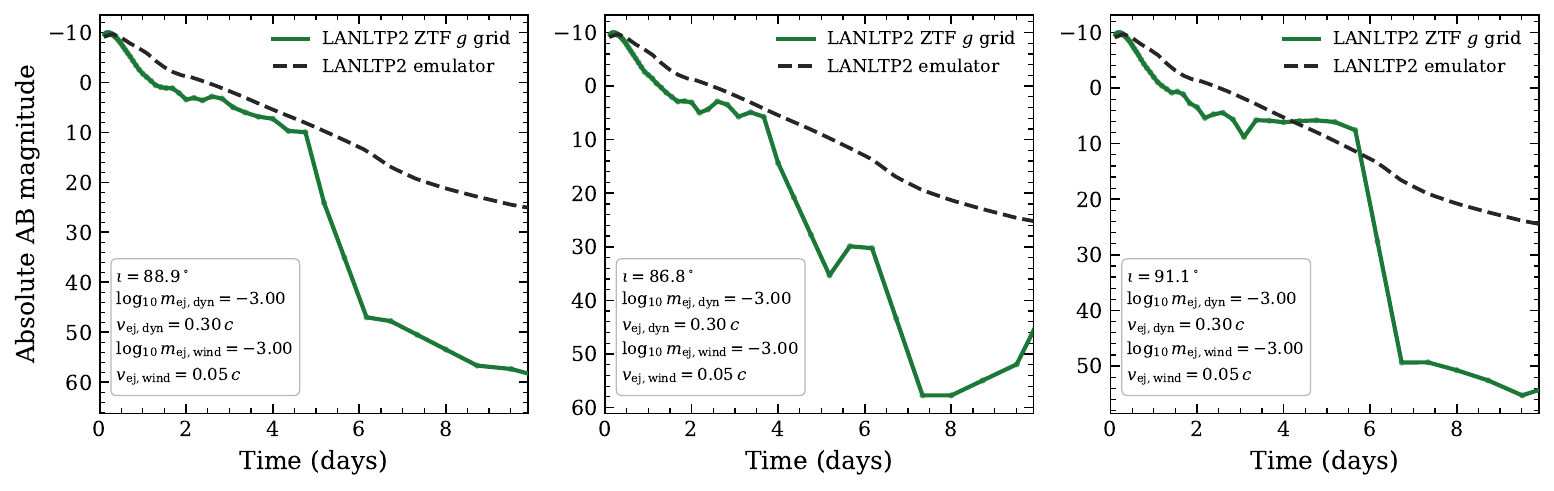}
  \caption{
  Surrogate-predictions (dashed) and training grid points (solid) absolute-magnitude light curves for the three highest mean MSE training grid points, all near ($\iota=90^\circ$). Panel titles report the viewing angle and ejecta configurations for each case. These non-smooth light curve shapes near the equatorial viewing angle are the origin of the high-MSE region identified in Figure~\ref{fig:val_theta}a.
  }
  \label{fig:appendix_high_mse}
\end{figure}

Figure~\ref{fig:val_theta} in Section~\ref{sec:photometry_validation} shows that surrogate accuracy degrades sharply near \(\iota\approx90^\circ\). Figure~\ref{fig:appendix_high_mse} shows the light curves predicted with the emulator for the three training grid points with the largest mean MSE in the ZTF \textit{g} band. The low mass ejecta grid points with $\iota\approx90^\circ\ $ are the culprit, which is also consistent with the intrinsic parameters of the highest MSE validation points. As seen in Figure~\ref{fig:appendix_high_mse} the grid points used for training have nonphysical features, specifically sharp and rapid declines, in their light curves for this regime, which leads to the high MSE. This is caused we approach the faintest edge of the parameter space, and the simulations we rely on lose fidelity. The emulator predictions do however remain more smooth in agreement with the rest of the parameter space. Regardless, it is unlikely to observe an off-axis, extremely low mass ejecta kilonova, as it will be very faint. 
%, and Table~\ref{tab:appendix_high_mse} lists their ejecta parameters. %All eight lie within \(10^\circ\) of the equator (\(\iota\approx87^\circ\)--\(95^\circ\)), a far higher rate than expected if MSE were independent of viewing angle (\(8/8\) observed vs.\ \(\approx0.9/8\) expected for a uniform draw over the sampled \(\iota\) range).
All three lie within $5^\circ$ of the equator ($\iota\approx86^\circ$--$92^\circ$). Under a uniform, angle-independent null, only $8\times(20/180)\approx0.9$ of the eight worst-fitting nodes would be expected in this window by chance.
The predicted and true light curves in this region show non-smooth, irregular structure that the flux-based surrogate does not reproduce, which is the direct source of the elevated MSE.

%The ejecta masses of these eight configurations are only modestly displaced from the full 66-configuration sample: the mean \(\log_{10}(M_{\rm ej}^{\rm dyn}/M_\odot)\) is \(-2.05\) among the eight worst-fitting cases versus \(-1.92\) over the full sample, while the mean
%\(\log_{10}(M_{\rm ej}^{\rm wind}/M_\odot)\) is essentially unchanged (\(-1.71\) vs.\ \(-1.71\)). This is consistent with a modest contribution from low dynamical-ejecta mass compounding the viewing-angle effect, but with only eight configurations we do not treat this as a statistically robust trend; the dominant, clearly resolved effect is the clustering in \(\iota\).

%\begin{table}[h]
%\centering
%\caption{Ejecta parameters of the eight validation configurations with the highest mean MSE
%(Figure~\ref{fig:appendix_high_mse}), ranked worst (rank 0) to eighth-worst.}
%\label{tab:appendix_high_mse}
%\begin{tabular}{ccccccc}
%\hline
%ank & $\iota$ (deg) & MSE (mag$^2$) & $\log_{10}(M_{\rm ej}^{\rm dyn}/M_\odot)$ & $v_{\rm ej}^{\rm dyn}/c$ & $\log_{10}(M_{\rm ej}^{\rm wind}/M_\odot)$ & $v_{\rm ej}^{\rm wind}/c$ \\
%\hline
%0 & 88.9 & 53.2 & $-2.77$ & 0.181 & $-2.94$ & 0.058 \\
%1 & 88.9 & 32.1 & $-2.22$ & 0.196 & $-1.67$ & 0.055 \\
%2 & 91.1 & 29.0 & $-2.22$ & 0.196 & $-1.67$ & 0.055 \\
%3 & 91.1 & 29.0 & $-2.77$ & 0.181 & $-2.94$ & 0.058 \\
%4 & 95.4 & 25.7 & $-1.20$ & 0.293 & $-1.09$ & 0.057 \\
%5 & 93.2 & 25.4 & $-1.20$ & 0.293 & $-1.09$ & 0.057 \\
%6 & 86.8 & 23.7 & $-1.99$ & 0.287 & $-1.13$ & 0.076 \\
%7 & 88.9 & 21.0 & $-1.99$ & 0.287 & $-1.13$ & 0.076 \\
%\hline
%\end{tabular}
%\end{table}

%\begin{figure}[t]
  %\centering
  %\includegraphics[width=\columnwidth]{figures/fig_appendix_high_mse.pdf}
  %\caption{
  %Surrogate-predicted (solid) and \texttt{SuperNu} (dashed) absolute-magnitude light curves for the eight validation configurations with the highest mean MSE (Table~\ref{tab:appendix_high_mse}), all within \(10^\circ\) of \(\iota=90^\circ\). Panel titles report the viewing angle and mean MSE for each case. These non-smooth light curve shapes near the equatorial viewing angle are the origin of the high-MSE region identified in Figure~\ref{fig:val_theta}a.
  %}
  %\label{fig:appendix_high_mse}
%\end{figure}

\bibliography{bib.bib}
\bibliographystyle{aasjournalv7}



\end{document}